\documentclass[aps,prd,twocolumn,groupedaddress,nofootinbib,amssymb,eqsecnum,epsfig]{revtex4}
\usepackage{graphicx}
\usepackage{bm}
\usepackage{dcolumn}
\usepackage{amsmath}

\def \lleq {\lower0.9ex\hbox{ $\buildrel < \over \sim$} ~}
\def \ggeq {\lower0.9ex\hbox{ $\buildrel > \over \sim$} ~}

\def \lcdm    {$\Lambda$CDM }

\def \omms   {\Omega_{\rm m}}

\def \beq  {\begin{equation}}
\def \eeq  {\end{equation}}
\def \ber  {\begin{eqnarray}}
\def \eer  {\end{eqnarray}}
\def \Geff {G_{\rm eff}}
\def \dotGeff {\dot{G}_{\rm eff}}
\def \ddotGeff {\ddot{G}_{\rm eff}}
\def \ommgr {\Omega_{\rm m,GR}}
\newcommand{\fs}{{\rm{\it fs}}8}

\begin{document}
\newcommand{\newc}{\newcommand}

\newc{\be}{\begin{equation}}
\newc{\ee}{\end{equation}}
\newc{\ba}{\begin{eqnarray}}
\newc{\ea}{\end{eqnarray}}
\newc{\bea}{\begin{eqnarray*}}
\newc{\eea}{\end{eqnarray*}}
\newc{\D}{\partial}
\newc{\ie}{{\it i.e.} }
\newc{\eg}{{\it e.g.} }
\newc{\etc}{{\it etc.} }
\newc{\etal}{{\it et al.}}
\newcommand{\nn}{\nonumber}
\newc{\ra}{\rightarrow}
\newc{\lra}{\leftrightarrow}
\newc{\lsim}{\buildrel{<}\over{\sim}}
\newc{\gsim}{\buildrel{>}\over{\sim}}
\title{The WiggleZ Dark Energy Survey: constraining the evolution of Newton's constant using the growth rate of structure}

\author{Savvas Nesseris$^{1}$, Chris Blake$^2$, Tamara Davis$^{3,4}$, David Parkinson$^3$}
%
%
\affiliation{$^{1}$ The Niels Bohr International Academy and Discovery Center, The Niels Bohr Institute, Blegdamsvej 17,
DK-2100, Copenhagen \O, Denmark \\ $^2$
Centre for Astrophysics \& Supercomputing, Swinburne University of
Technology, P.O. Box 218, Hawthorn, VIC 3122, Australia \\ $^3$ School
of Mathematics and Physics, University of Queensland, Brisbane, QLD
4072, Australia \\ $^4$ Dark Cosmology Centre, Niels Bohr Institute,
University of Copenhagen, Juliane Maries Vej 30, DK-2100 Copenhagen \O, Denmark}

\date{\today}

\begin{abstract}
We constrain the
evolution of Newton's constant using the growth rate of large-scale
structure measured by the WiggleZ Dark Energy Survey in the redshift
range $0.1 < z < 0.9$.  We use this data in two ways. Firstly we constrain the matter density of the Universe, $\omms$ (assuming General
Relativity), and use this to construct a diagnostic to detect the presence of an
evolving Newton's constant. Secondly we directly measure the evolution of Newton's
constant, $\Geff$, that
appears in Modified Gravity theories, without assuming General Relativity to be
true. The novelty of these approaches are that, contrary to other methods, they do
not require knowledge of the expansion history of the Universe, $H(z)$, making them
model independent tests. Our constraints for the second derivative of Newton's constant at the present day, assuming it is slowly evolving as suggested by Big Bang Nucleosynthesis constraints, using the WiggleZ data is $\ddotGeff(t_0)=-1.19\pm 0.95\cdot 10^{-20}h^2 \textrm{yr}^{-2}$, where $h$ is defined via $H_0=100~h~km~ s^{-1}Mpc^{-1}$, while using both the WiggleZ and the Sloan Digital Sky Survey Luminous Red Galaxy (SDSS LRG) data is $\ddotGeff(t_0)=-3.6\pm 6.8\cdot 10^{-21}h^2 \rm{yr}^{-2}$, both being consistent with General Relativity. Finally, our constraint for the rms mass fluctuation $\sigma_8$ using the WiggleZ data is $\sigma_8=0.75 \pm 0.08$, while using both the WiggleZ and the SDSS LRG data $\sigma_8=0.77 \pm 0.07$, both in good agreement with the latest measurements from the Cosmic Microwave Background radiation.
\end{abstract}

\maketitle

\section{Introduction}
Recent observations suggest that the Universe is undergoing a phase of accelerated expansion, usually attributed to an unknown ideal fluid dubbed dark energy.
Most of the evidence for the existence of dark energy comes from geometric tests that measure the expansion rate of the universe
$H(a)\equiv \dot{a}/a$ at different epochs, where $a(t)$ is the scale factor of a Friedmann-Robertson-Walker metric. Examples of such tests include measurements of the luminosity distance, $d_L(a)$, using standard candles like Type Ia supernovae (SN Ia) \cite{Kilbinger:2008gk, Kessler:2009ys} and measurements of the angular diameter distance, $d_A(a)$, using standard rulers such as the scale of the sound horizon at last scattering \cite{Komatsu:2010fb} and baryon acoustic oscillations \cite{Percival:2009xn}. Even though these tests are presently
the most accurate probes of dark energy, the mere determination of the expansion rate, $H(a)$, is not able to provide significant insight into the properties of dark energy
to distinguish it from models that attribute the accelerating expansion to modifications of general relativity \cite{Tsujikawa:2010zza,Heavens:2007ka,Nesseris:2006er,Nesseris:2006jc}.

A smoking gun signature of modified gravity theories, such as $f(R)$ gravity  \cite{fRpapers} or scalar-tensor models \cite{stensor}, is the fact that they predict that Newton's constant $G_N$ is evolving with time. Generically, modified gravity theories can be cast in such a way as to have an effective Newton's constant that changes with time $\Geff(t)$ \cite{DeFelice:2010gb, Tsujikawa:2007gd, DeFelice:2010aj}. However, this variation cannot be arbitrarily large and has already been constrained by Big Bang Nucleosynthesis to be within $10\%$.  See, for example, Ref.~\cite{Bambi:2005fi} which gives $\Geff(t_{BBN})/\Geff(t_0)=1.09\pm^{0.22}_{0.19}$, where $t_{\rm BBN}$ is the time of Big Bang Nucleosynthesis, while $t_0$ is the present day.  This allows us to make the simplifying assumption that $\Geff$ is a slowly varying function. Furthermore, Solar System tests and SN Ia observations (see \cite{Nesseris:2006hp} and references therein) suggest that the first time derivative $\dotGeff(t_0)$ is almost zero and that for the second time derivative we have $|g_2|\lesssim O(1)$ , where $ g_2 \equiv \ddotGeff(t_0)/(\Geff(t_0)H_0^2)$.

The additional observational input that is required to discriminate between dark energy and modified gravity theories is the growth function of the linear matter density contrast $\delta\equiv\frac{\delta\rho}{\rho}$, where $\rho$ represents the background matter density and $\delta\rho$ its first order perturbation. The reason for this is the fact that the growth of cosmic structure is a result of the motion of matter and therefore is sensitive to both the expansion of the Universe $H(a)$ and $\Geff$.

More specifically, it can be shown that in modified gravity theories on sub-horizon scales, ie $k^2\gg a^2H^2$ where $k$ is the wave-number of the modes of the perturbations in Fourier space, the growth factor satisfies the following differential equation \cite{DeFelice:2010gb, Tsujikawa:2007gd, DeFelice:2010aj, Nesseris:2009jf}:
\be
\delta''(a)+\left(\frac{3}{a}+\frac{H'(a)}{H(a)}\right)\delta'(a)
-\frac{3}{2}\frac{\omms \Geff(a)}{a^5 H(a)^2/H_0^2}~\delta(a)=0
\label{ode}\ee where primes denote differentiation with respect to the scale factor,  $H(a)\equiv\frac{\dot{a}}{a}$ is the Hubble parameter and we assume the initial conditions $\delta(0)=0$ and $\delta'(0)=1$ for the growing mode. When $\Geff(a)=1$ we get GR as a subcase, while in general in modified gravity theories $\Geff(a)$ is time and scale dependent. For example, in $f(R)$ theories we have that \cite{Tsujikawa:2007gd} \ba \Geff&=&\frac{1}{8\pi F}\frac{1+4\frac{k^2}{a^2}m}{1+3\frac{k^2}{a^2}m} \label{gefffr1}\\ m&\equiv& \frac{F_{,R}}{F}\\F&\equiv&f_{,R}=\frac{\partial f}{\partial R}\label{gefffr2}\ea which reduces to GR only when $f(R)=R-2\Lambda$.

The exact solution of Eq.~(\ref{ode}) for a flat GR model with a constant dark energy equation of state $w$  is given for the growing mode by \cite{Silveira:1994yq,Percival:2005vm}:\ba \delta(a)&=& a \cdot {}_2F_1 \left(- \frac{1}{3 w},\frac{1}{2} -
\frac{1}{2 w};1 - \frac{5}{6 w};a^{-3 w}(1 - \omms^{-1})\right)
\label{Da1} \nn \\ \textrm{for}&&H(a)^2/H_0^2= \omms a^{-3}+(1-\omms)a^{-3(1+w)},\ea
where ${}_2F_1(a,b;c;z)$ is a hypergeometric function defined by the series
\be {}_2F_1(a,b;c;z)\equiv \frac{\Gamma(c)}{\Gamma(a)\Gamma(b)}\sum^{\infty}_{n=0}\frac{\Gamma(a+n)\Gamma(b+n)}{\Gamma(c+n)n!}z^n \ee on the disk $|z|<1$ and by analytic continuation elsewhere, see Ref.~\cite{handbook} for more details.

In more general cases it is not possible to find an analytical solution for Eq.~(\ref{ode}), but in Ref.~\cite{Wang:1998gt} it was shown that the growth rate $f(a)\equiv \frac{d ln \delta}{dlna}$ can be approximated as \ba f(a)=\omms(a)^\gamma \label{fg}\\\omms(a)\equiv\frac{\omms~a^{-3}}{H(a)^2/H_0^2} \\
\gamma \simeq \frac{3 (1-w)}{5-6 w} \ea where this approximation for $\gamma$ is  valid at first order for a dark energy model with a constant equation of state $w$ and for \lcdm we have $\gamma=\frac{6}{11}\simeq 0.55$. Conveniently, the approximation of Eq.~(\ref{fg}) can be used to accurately fit a wide variety of different scenarios, including modified gravity models, if $\gamma$ is allowed to be a free parameter.

In order to get a prediction for $\Geff$ from the growth rate data one would need to solve Eq.~(\ref{ode}) and this can only be done for a specific theory, as knowledge of $H(a)$ is required. Therefore, in order to reconstruct $\Geff$ the usual procedure would be to assume a model for $H(a)$, fit the approximation of Eq.~(\ref{fg}) to the data and finally use Eq.~(\ref{ode}) to get a prediction for $\Geff$. This obviously has several drawbacks, as the growth data alone are not sufficient to get viable fits to $H(a)$ and are usually complemented by SN Ia and CMB data. The SN Ia data actually determine the luminosity distance $d_L(a)$, therefore converting
these measurements to H(a) depends on differentiation of noisy data and an assumption about the spatial curvature $\Omega_K$ and this obviously results in large error bars for the best fit parameters. Finally, any result would ultimately depend on the specific choice of $H(a)$, or in other words it would suffer from model-choice bias, whereas what we propose is a model independent technique that can be used even without the CMB and SN Ia data. In order to solve these problems we constructed a diagnostic that is based solely on data, with no assumptions about the expansion history of the Universe, that is able to detect the presence of an evolving Newton's constant. We describe both our new method and the diagnostic in detail in Section \ref{method}.

\section{The WiggleZ data}
The WiggleZ Dark Energy Survey at the Anglo-Australian Telescope
\cite{Drinkwater:2009sd} is a large-scale galaxy redshift survey of
bright emission-line galaxies mapping a cosmic volume of order 1
Gpc$^3$ over the redshift interval $z < 1$.  The survey has obtained
of order $200{,}000$ redshifts for UV-selected galaxies covering of
order 1000 deg$^2$ of equatorial sky.  In this paper we analyze the
subset of the WiggleZ sample assembled up to the end of the 10A
semester (May 2010).  We include data from six survey regions in the
redshift range $0.1 < z < 0.9$ -- the 9-hr, 11-hr, 15-hr, 22-hr, 1-hr
and 3-hr regions -- which together constitute a total sample of $N =
152{,}117$ galaxies.

In Ref.~\cite{Blake:2011rj} the authors presented fits for the growth rate of structure
within this sample using redshift-space distortions in the 2D power
spectrum.  These fits included a full exploration of the systematic
errors arising from the assumption of redshift-space distortion models
based on perturbation theory techniques, fitting formulae calibrated by
N-body simulations, and empirical models.  The WiggleZ data were
analyzed in four redshift slices with effective redshifts $z = (0.22,
0.41, 0.6, 0.78)$.  The growth rate measurements determine the quantity
$f(z) \, \sigma_8(z) = (0.42 \pm 0.07, 0.45 \pm 0.04, 0.43 \pm 0.04,
0.38 \pm 0.04)$ at these four redshifts.  The reader is referred to
Ref.~\cite{Blake:2011rj} for full details of how these measurements were
performed.

\section{The method}\label{method}
Instead of solving Eq.~(\ref{ode}) for $\delta(a)$, we will solve it for $H(a)$ \cite{Starobinsky:1998fr, Nesseris:2007pa}.
Then we get \footnote{See Appendix \ref{apHa} for more details.}
\be H(a)^2/H_0^2= \frac{3\omms}{a^6 \delta'(a)^2} \int^a_0 x ~ \Geff(x)
\delta(x)\delta'(x)dx   \label{Ha} \ee

The growth rate is defined as $f(a)\equiv \frac{d ln \delta}{dlna}$ and therefore
$\delta$ can be written in terms of $f$ as $\delta=\delta(1) e^{\int_1^a f(x)/x dx}$.
However, WiggleZ gives the data in the form $f \cdot \sigma_8$, where the rms mass
fluctuation $\sigma_8(z)$ can be expressed as \cite{Wang:1998gt} $\sigma_8(z)=\frac{\sigma_8(1)}{\delta(1)}\delta(z)$
and as a result we can express the product as
$f(a) \cdot \sigma_8(a)=\frac{\sigma_8(1)}{\delta(1)}~a~\delta'(a)$. So, it would be
convenient to define a new parameter \be \fs(a)\equiv f(a) \cdot \sigma_8(a)=\xi a \delta '(a)\label{fs8}\ee
where $\xi \equiv \frac{\sigma_8(1)}{\delta(1)}$. Using this we get for the growth
factor \be \delta(a)=\int_0^a \frac{\fs(x)}{\xi x} dx \label{growth}\ee while using Eqs.~(\ref{Ha}) and (\ref{growth}) we get for the Hubble parameter $H(a)$ \be H(a)^2/H_0^2=\frac{3\omms}{a^4 \fs(a)^2} \int_0^a dx~\fs(x) \Geff(x) \int_0^x dy \frac{\fs(y)}{y} \label{Ha1}\ee The parameter $\xi$ has canceled out, so evaluating Eq.~(\ref{Ha1}) at $a=1$ and solving for $\omms$ we get \ba \frac{1}{3\omms}&=&\int_0^1 dx~\Geff(x)~\frac{\fs(x)}{\fs(1)} \int_0^x dy\frac{1}{y} \frac{\fs(y)}{\fs(1)} \nn \\ &=& \int_0^1 dx~\Geff(x)~g(x) \label{om}\ea where \be g(x)\equiv ~\frac{\fs(x)}{\fs(1)}~\int_0^x dy \frac{1}{y}\frac{\fs(y)}{\fs(1)} \ee Notice that Eq.~(\ref{om}) does not include the Hubble parameter $H(z)$ or any assumption of some dark energy model at all. Therefore, the calculation of $\omms$ or $\Geff$ can be carried out by using only the growth rate data. Now, Taylor expanding $\Geff$ around $a=1$ \be \Geff(a)=\sum_{n=0}^{\infty}g_n \frac{(a-1)^n}{n!}\ee where we have set $g_n=\frac{d^n \Geff(a)}{da^n}|_{a=1}$, and plugging it into Eq.~(\ref{om}) we get \ba \frac{1}{3\omms}&=&\int_0^1 dx~\sum_{n=0}^{\infty}g_n \frac{(x-1)^n}{n!} g(x) \nn \\ &=&\sum_{n=0}^{\infty}g_n~\int_0^1 dx~g(x) \frac{(x-1)^n}{n!} \nn \\ &=& \sum_{n=0}^{\infty}g_n I_n \label{con1}\ea where \be I_n \equiv \int_0^1 dx~g(x) \frac{(x-1)^n}{n!}. \ee Clearly, the constants $I_n$ depend solely on the data.
We can get another constraint by demanding that $\Geff(a=0)=1$, then using the Taylor expansion we get \be \sum_{n=0}^{\infty}g_n \frac{(-1)^n}{n!}=1\label{con2}\ee If we don't make the assumption that $\Geff(a\ll1)\rightarrow1$ then it can be shown that the growth of perturbations $\delta(a)$ in a matter dominated Universe ($\omms=1$), but with a non-constant $\Geff$, does not scale like $\delta(a)\sim a$ for $a\ll1$ as expected\footnote{See Appendix \ref{formulas} for analytic formulas for $\delta(a)$ in a matter dominated Universe with a non-constant $\Geff$.}. Also, since modified gravity models are used as alternative explanations of dark energy, which is prevalent at late times, any deviations from GR would also be expected at late times not at early times. We can use Eq.~(\ref{con2}) to eliminate $g_0$, since \be g_0=1-\sum_{n=1}^{\infty}g_n\frac{(-1)^n}{n!} \ee and from Eq.~(\ref{con1}) we get \be g_0 I_0 +\sum_{n=1}^{\infty} g_n I_n=\frac{1}{3\omms} \label{sum1}\ee or by combining these two and rearranging \be \sum_{n=1}^{\infty}g_n \left( I_n-\frac{(-1)^n}{n!} I_0\right)=\frac{1}{3\omms}-I_0 .\ee

In GR we have $g_0=1$ and $g_n=0$ for $n\geq1$, so we can define the GR value of $\omms$ as
\be \frac{1}{3\Omega_{\rm m,GR}} \equiv \int_0^1 dx~g(x)=I_0 \label{omGR}\ee while if we could determine the form of $\Geff(a)$ accurately then we would define the ``real'' value of $\omms$ as
\ba \frac{1}{3\Omega_{\rm m,real}} &\equiv& \int_0^1 dx~\Geff(x)~g(x)=\nn \\&=&I_0+\sum_{n=1}^{\infty}g_n \left( I_n-\frac{(-1)^n}{n!} I_0\right). \label{omreal}\ea So, if we form the ratio between the two parameters we have \ba r&\equiv&\frac{\Omega_{\rm m,GR}}{\Omega_{\rm m,real}}=\frac{\int_0^1 dx~\Geff(x)~g(x)}{\int_0^1 dx~g(x)} =\nn \\ &=& 1+\sum_{n=1}^{\infty}g_n \left( \frac{I_n}{I_0}-\frac{(-1)^n}{n!}\right).\label{ratio}\ea Also, Eq.~(\ref{omGR}) can be used to construct a diagnostic \be {\O}\equiv \frac{\omms-\Omega_{\rm m,GR}}{\omms}=1-\frac{\Omega_{\rm m,GR}}{\omms}=1-\frac{1}{3 I_0 \omms}\label{diag}\ee where $\omms$ is the value of the matter density as measured from other independent observations. Therefore, if we assume that the value of $\omms$ is independently and accurately determined by other observations, then any statistically significant deviation of the quantity ${\O}$ from zero or the ratio $r$ from $1$, clearly and uniquely identifies an evolving $\Geff$ and consequently favours modified gravity theories. Finally, notice that the diagnostic ${\O}$ depends solely on measurable quantities and not on any model.

For concreteness we will also consider a specific model for $\Geff$. However, any variation of Newton's constant can be constrained by BBN to be less than $~10\%$, see for example Ref.~\cite{Bambi:2005fi} who report $\Geff(t_{\rm BBN})/\Geff(t_0)=1.09\pm^{0.22}_{0.19}$, which motivates us to assume that $\Geff$ is a slowly varying function. Also, the Solar System tests and SN Ia observations (see \cite{Nesseris:2006hp} and references therein) suggest that $|g_1| \simeq10^{-3} h^{-1} \ll 1$ and $|g_2| \sim O(1)$, so we can safely neglect $g_1$. However, the measurable quantities are $\dotGeff(t_0)$ and $\ddotGeff(t_0)$, so we should express these (after a bit of algebra) as $\dotGeff(t_0)=g_1 H_0$ and $\ddotGeff(t_0)=H_0^2\left(g_1(1+\frac{H'(1)}{H_0})+g_2 \right)$. So, after neglecting $g_1$, the expression for the second derivative can be simplified to $\ddotGeff(t_0)\simeq g_2 H_0^2$. Furthermore, modified gravity models are used as alternative explanations of dark energy, which is prevalent at late times, so any deviations from GR would be expected at late times not early times. As a result we will demand that $\Geff(a \ll 1)\rightarrow1$.

Additionally, we will not consider models with wildly oscillating $\Geff$. The reason for this is that since $\Geff$ is calculated via differentiations of the gravity Lagrangian, for example in the $f(R)$ case see Eqs. (\ref{gefffr1}) - (\ref{gefffr2}), then one would need to have a Lagrangian with terms that exhibit explicit oscillatory behavior. In the case of the $f(R)$ theories one could consider the Lagrangian $f(R)=R+A~\sin(B~R)$, where $A$, $B$ are constants that play the role of the ``amplitude" and the ``frequency" of the oscillations respectively. However, it is far from trivial to construct and (more importantly) to theoretically motivate such a model, and in any case, most viable $f(R)$ or other modified gravity models in the literature predict a smooth and very slowly evolving $\Geff$.

This is the motivation for our decision to consider only smooth and slowly evolving models for $\Geff$. Therefore, our assumptions for the specific model for $\Geff$ are the following:
\begin{enumerate}
  \item We assume $\Geff(a)$ has a zero first derivative at $a=1$ and is slowly varying.
  \item We assume that $\Geff(a\ll1)\rightarrow1$, because dark energy should be negligible at early times.

\end{enumerate}
The second point implies that the present day Newton's constant has not been scaled to 1 in our conventions and as a result, from the moment we chose the normalization, any deviation of $\Geff$ from 1 is a GR violation. Under these assumptions, we can approximate $\Geff$ by a series expansion as follows \ba \Geff(a) &=& \Geff(1)+\frac{1}{2}\Geff''(1)(a-1)^2 \nn \\&=&
g_0+\frac{1}{2}g_2(a-1)^2 \ea where we have set $g_0=\Geff(1)$ and $g_2=\Geff''(1)$. Note that the case $(g_0,g_2)=(1,0)$ corresponds to GR. Under the assumption $\Geff(a\ll1)\rightarrow1$ we can fix $g_0=1-\frac{g_2}{2}$ and as a result we have, \be \Geff(a) = 1-\frac{g_2}{2}+\frac{1}{2}g_2(a-1)^2 \label{geff}\ee For the choice of Eq.~(\ref{geff}) we can see that the ratio $r=\frac{\Omega_{\rm m,GR}}{\Omega_{\rm m,real}}$ can be written as: \ba r&=&1+g_2 \left(\frac{I_2}{I_0}-\frac{1}{2}\right)=1+g_2\frac{2 I_2-I_0}{2I_0}\nn\\ &=&1-g_2 \frac{1}{2I_0}\int_0^1 dx~g(x)\left(1-(x-1)^2\right).\label{ratiog2} \ea Whether the ratio is larger or smaller than $1$ depends on both the sign of $g_2$ and the sign of the integral in the last line of Eq.~(\ref{ratiog2}). Since $g(a)>0$ for $0<a<1$, we can see that both the constant $I_0$ and the integral in the last line of Eq.~(\ref{ratiog2}) will be positive. So, when $g_2<0$ the ratio $r$ of Eq.~(\ref{ratiog2}) will be larger than $1$, thus giving the impression that $\Omega_{\rm m,GR}$ is overestimated compared to the real value $\Omega_{\rm m,real}$ and vice versa when $g_2>0$.

Finally, using the fact that $g_1\ll1$ we can get an expression for $g_2$ given a value for $\omms$ \be g_2 =\frac{\frac{1}{3\omms}-I_0}{I_2-I_0/2}\label{g2}\ee where we of course assume, based on the slowly varying nature of $\Geff$, that for the explicit example of Eq.~(\ref{geff}) the higher derivatives $g_n$ for $n\geq3$ are much smaller than $g_2$.

Therefore, we have two options:
\begin{itemize}
\item we can use Eqs.~(\ref{omGR}) and (\ref{diag}) to calculate $\ommgr$ and the diagnostic $\O$ respectively,
\item or we can assume some value for $\omms$ and use Eq.~(\ref{g2}) to calculate $g_2$ and $\ddotGeff(t_0)$, under the assumption of a slowly varying $\Geff(a)$.
\end{itemize}
We will explore both of these two possibilities in the next section.

\section{Results}

\subsection{Tests and simulations}

\begin{table*}
\begin{center}
\caption{Results from the tests of our analysis method for simulated noise-free data for three different values of $g_2=(0,-1,1)$. In all cases the input model was $\omms=0.27$ and $\sigma_8=0.8$, while Case 1 corresponds to $(w_0,w_a)=(-0.95,0)$ and Case 2 to $(w_0,w_a)=(-0.95,0.5)$. These parameters were chosen in order to demonstrate the applicability of the method even in scenarios with an evolving dark energy equation of state $w(a)$. We then fit this data to four different models for the behavior of $\fs(a)$, and the numbers in brackets under each model show the  recovered values for $(\ommgr,g_2)$ while the first column labeled ``Original $g_2$" shows the assumed value of $g_2$ by which the data were constructed.} \label{table1}
\begin{tabular}{ccccccc}
\hline \vspace{-5pt}\\
   \vspace{5pt} \hspace{2pt} & \hspace{2pt}Original $g_2$& \hspace{2pt}Model 1  \hspace{2pt} & \hspace{2pt} Model 2  \hspace{2pt} & \hspace{2pt} Model 3  \hspace{2pt} & \hspace{2pt} Model 4 \\

\hline \hline \\

\vspace{6pt} \hspace{2pt}  & \hspace{2pt} 0    \hspace{2pt} & \hspace{5pt} (0.27,0.00) \hspace{2pt} & \hspace{2pt} (0.27,-0.01) \hspace{2pt} & \hspace{2pt} (0.27,0.00) \hspace{2pt} & \hspace{2pt} (0.24,0.27)\\

\vspace{6pt} Case 1 \hspace{2pt} & \hspace{2pt} -1 \hspace{2pt} & \hspace{5pt} (0.37,-0.86) \hspace{2pt} & \hspace{2pt} (0.37,-0.86) \hspace{2pt} & \hspace{2pt} (0.40,-1.08) \hspace{2pt} & \hspace{2pt} (0.37,-0.84)\\

\vspace{6pt}  \hspace{2pt} & \hspace{2pt} +1 \hspace{2pt} & \hspace{5pt} (0.17,0.92) \hspace{2pt} & \hspace{2pt} (0.17,0.91) \hspace{2pt} & \hspace{2pt} (0.16,1.02) \hspace{2pt} & \hspace{2pt} (0.11,1.39)\\ \hline \\

\vspace{6pt} \hspace{2pt} & \hspace{2pt} 0    \hspace{2pt} & \hspace{5pt} (0.28,-0.09) \hspace{2pt} & \hspace{2pt} (0.28,-0.10) \hspace{2pt} & \hspace{2pt} (0.27,0.00) \hspace{2pt} & \hspace{2pt} (0.25,0.22)\\

\vspace{6pt} Case 2 \hspace{2pt} & \hspace{2pt} -1 \hspace{2pt} & \hspace{5pt} (0.38,-0.98) \hspace{2pt} & \hspace{2pt} (0.38,-0.98) \hspace{2pt} & \hspace{2pt} (0.39,-1.04) \hspace{2pt} & \hspace{2pt} (0.38,-0.91)\\

\vspace{6pt}  \hspace{2pt} & \hspace{2pt} +1 \hspace{2pt} & \hspace{5pt} (0.17,0.88) \hspace{2pt} & \hspace{2pt} (0.17,0.87) \hspace{2pt} & \hspace{2pt} (0.16,0.99) \hspace{2pt} & \hspace{2pt} (0.12,1.35)\\
  \hline
\end{tabular}
\end{center}
\end{table*}

In order to examine the accuracy of our method we performed two tests, one with simulated noise-free data and one with simulated noisy data.

\begin{figure*}[t!]
\vspace{0cm}\rotatebox{0}{\vspace{0cm}\resizebox{.47\textwidth}{!}{\includegraphics{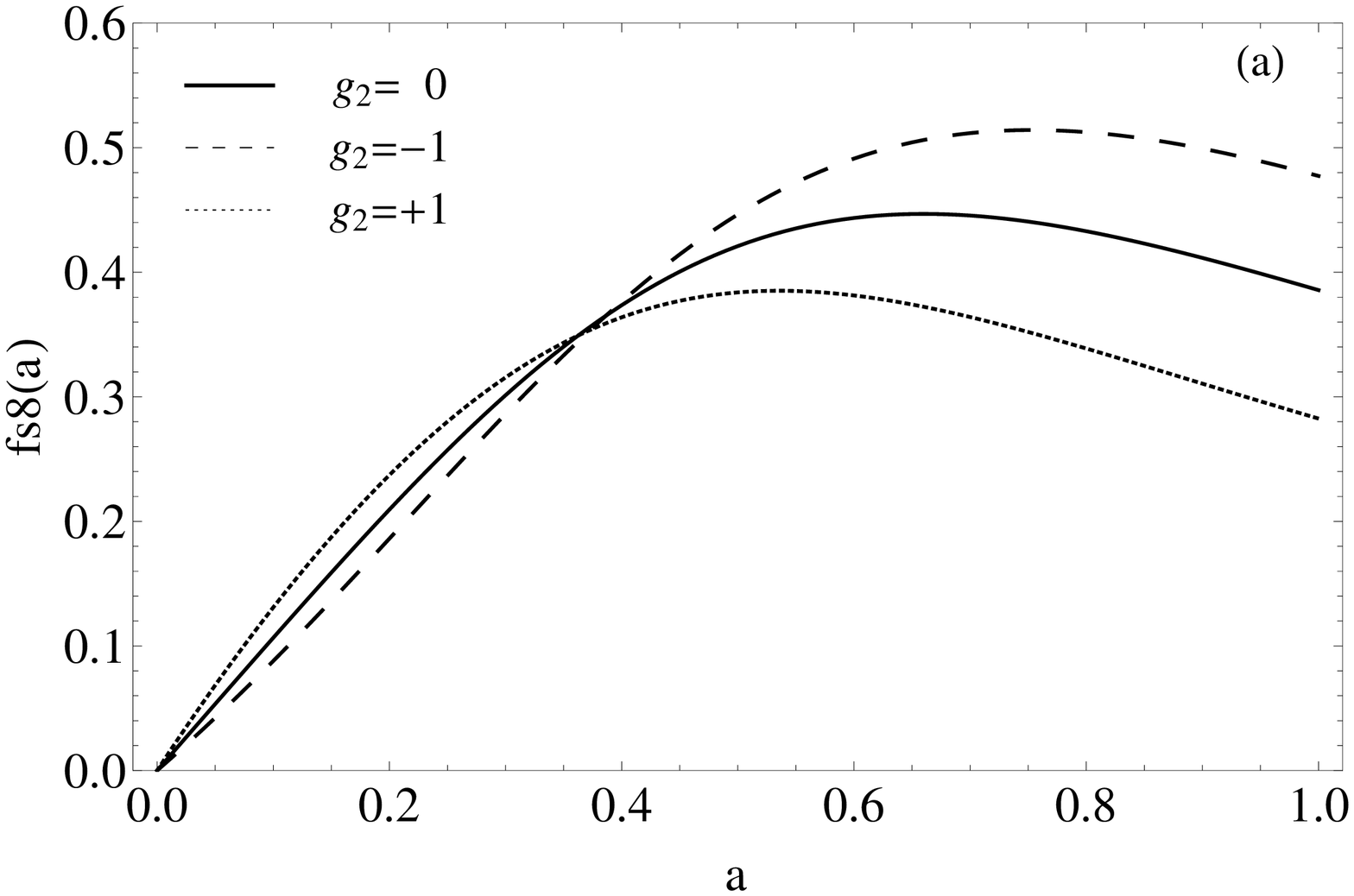}}}
\hspace{0.25cm}
\vspace{0cm}\rotatebox{0}{\vspace{0cm}\resizebox{.47\textwidth}{!}{\includegraphics{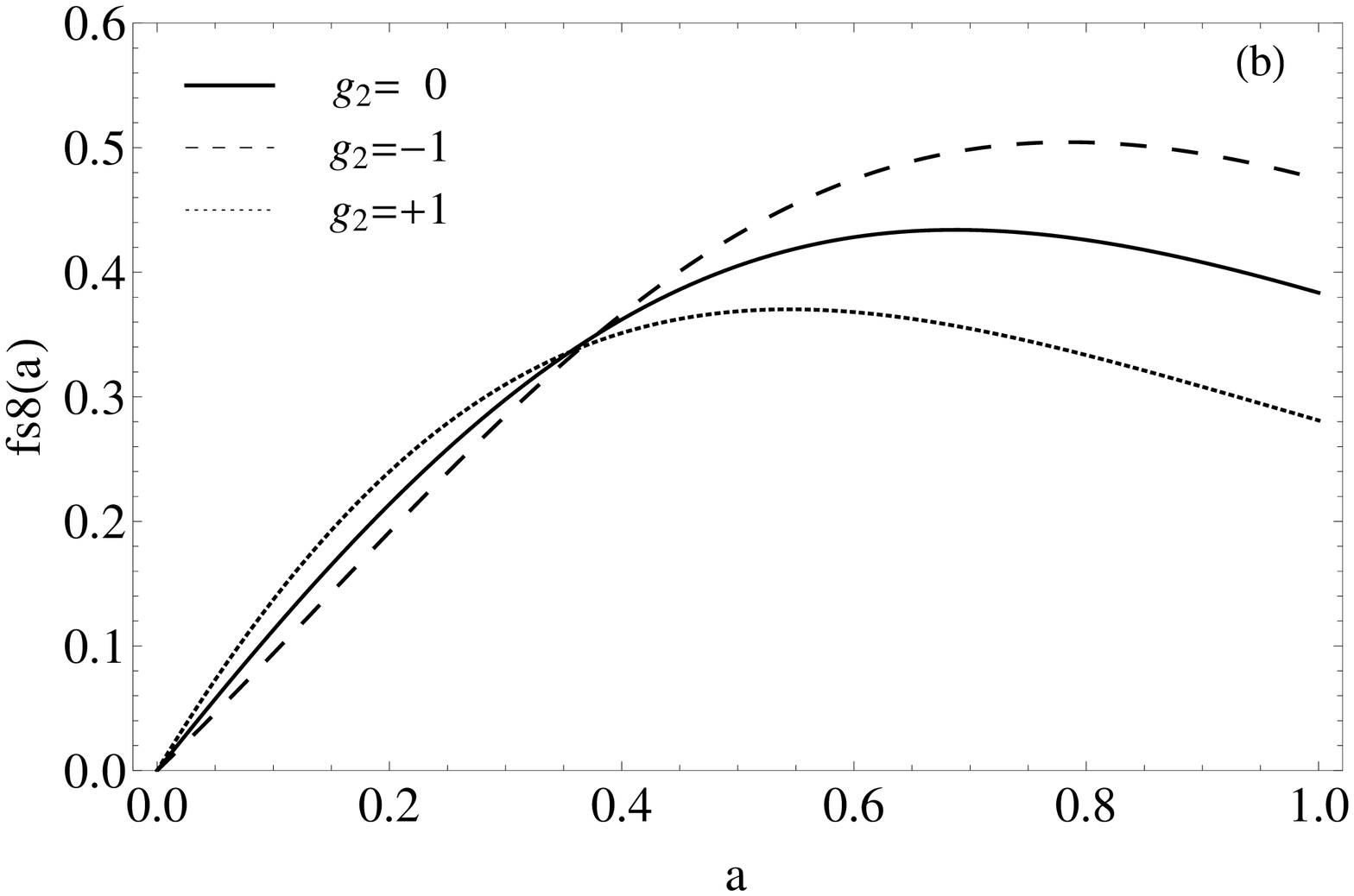}}}
\caption{The evolution of $\fs(a)$ for the two different cases considered in the text. Fig.~\ref{fig1}a shows Case 1 that has $(w_0,w_a)=(-0.95,0)$ and $g_2=(0,-1,1)$ while Fig.~\ref{fig1}b shows Case 2 that has $(w_0,w_a)=(-0.95,0.5)$. In both figures the values $g_2=(0,-1,1)$ correspond to the black solid, dashed and dotted lines respectively.\label{fig1}}
\vspace{0.5cm}
\vspace{0cm}\rotatebox{0}{\vspace{0cm}\resizebox{.47\textwidth}{!}{\includegraphics{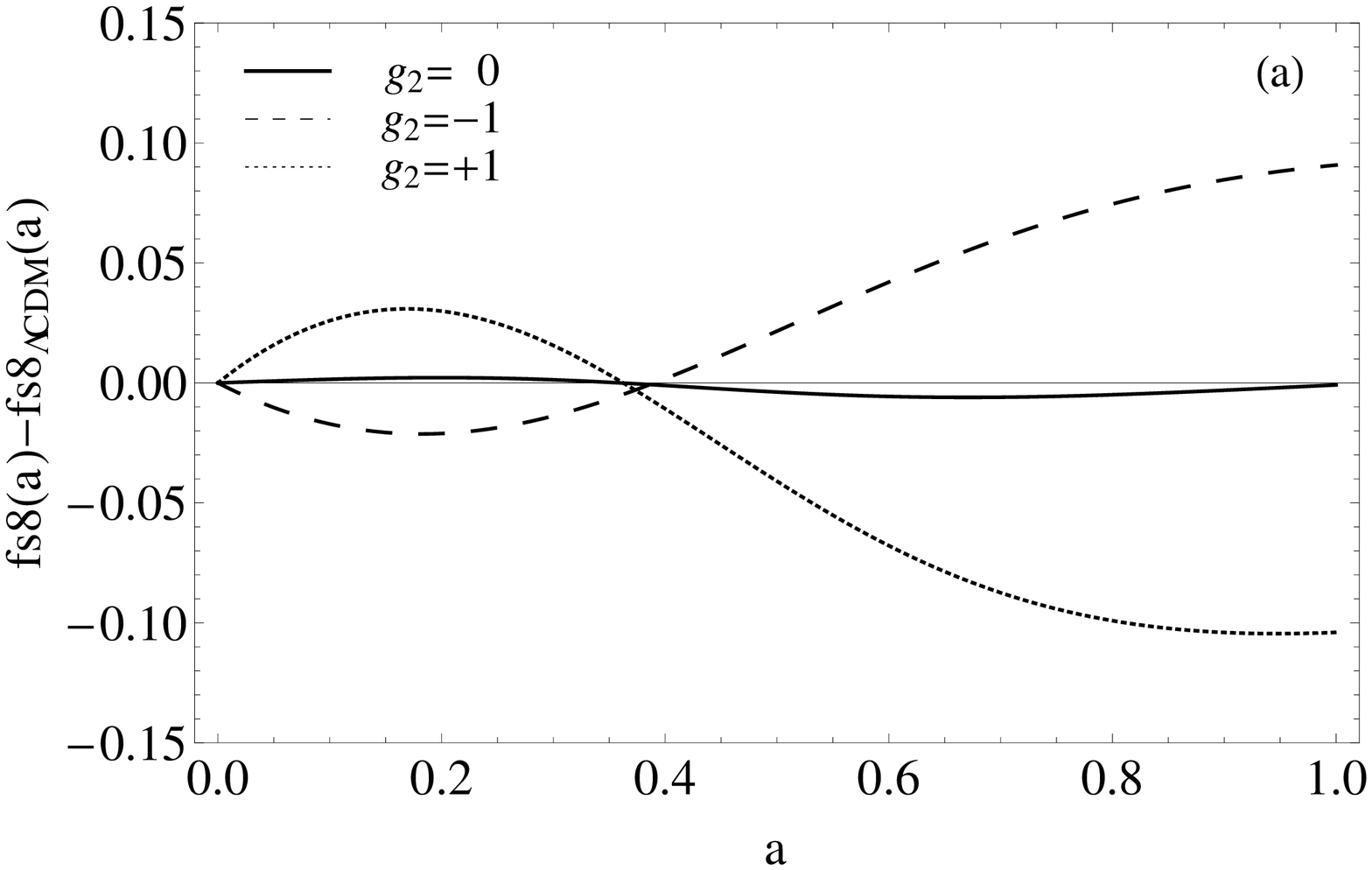}}}
\hspace{0.25cm}
\vspace{0cm}\rotatebox{0}{\vspace{0cm}\resizebox{.47\textwidth}{!}{\includegraphics{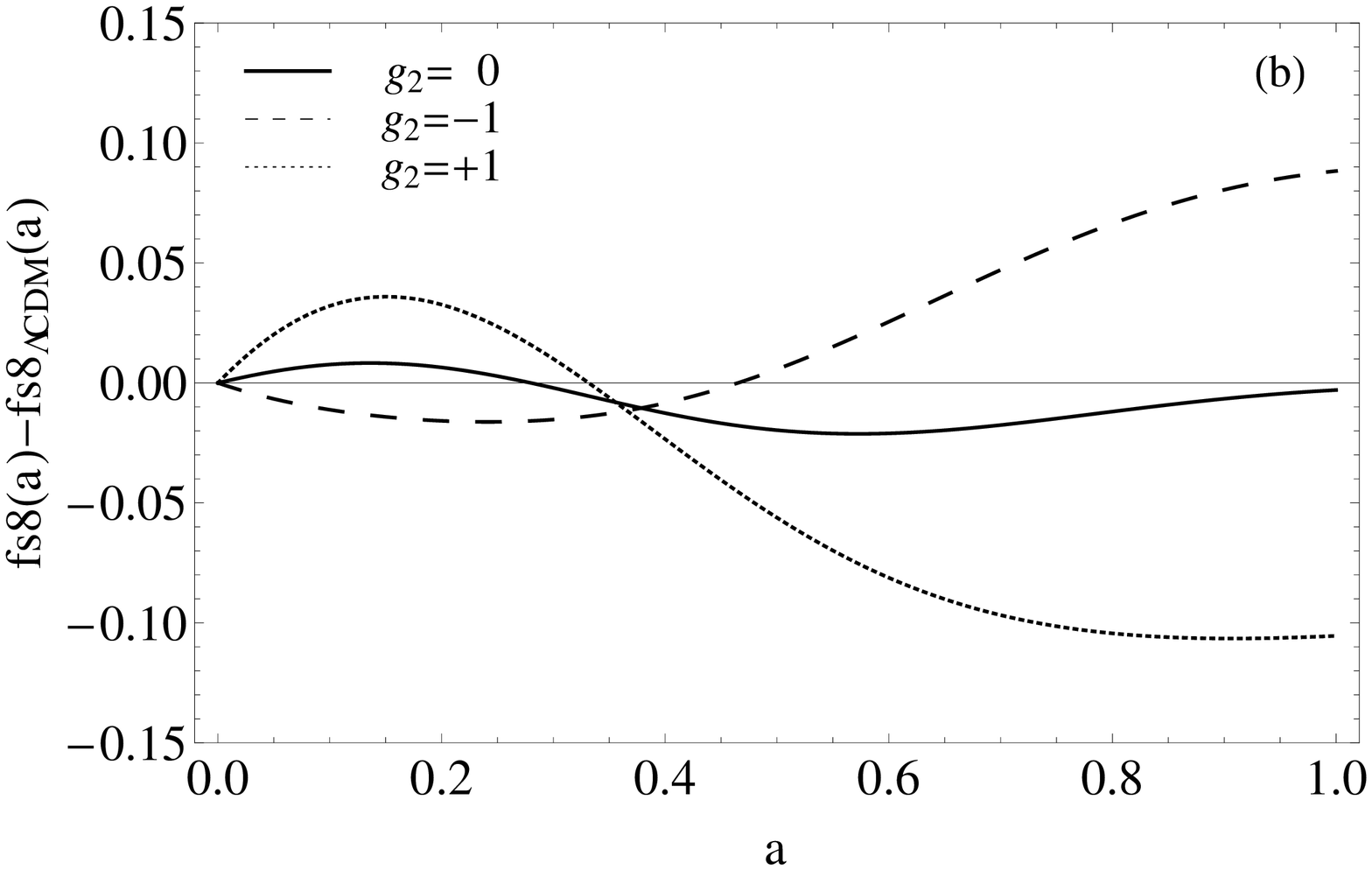}}}
\caption{The difference in the evolution of $\fs(a)$ between each case and the \lcdm model for $\omms=0.27$ and $\sigma_8=0.8$. Fig.~\ref{fig2}a shows Case 1 that has $(w_0,w_a)=(-0.95,0)$ and $g_2=(0,-1,1)$ while Fig.~\ref{fig2}b shows Case 2 that has $(w_0,w_a)=(-0.95,0.5)$. In both figures the values $g_2=(0,-1,1)$ correspond to the black solid, dashed and dotted lines respectively.\label{fig2}}
\end{figure*}

To generate the simulated noise-free data we numerically solved Eq.~(\ref{ode}) for six models, which we divide into two cases depending on whether they have a constant or varying equation of state of dark energy and whether they have a constant or varying gravitational parameter:

\begin{enumerate}
\item Case 1: Constant equation of state $w$ with parameters $(w_0,w_a)=(-0.95,0)$ and three values for the parameter $g_2=(-1,0,1)$
\item Case 2: Evolving equation of state $(w_0,w_a)=(-0.95,0.5)$ and three values for the parameter $g_2=(-1,0,1)$
\end{enumerate}
where the parameters $(w_0,w_a)$ correspond to a model for the dark energy equation of state $w(a)=w_0+w_a(1-a)$. We chose to use the values $(w_0,w_a)=(-0.95,0)$ and $(w_0,w_a)=(-0.95,0.5)$ in our simulations in order to test whether our method would be sensitive to models with a varying equation of state of dark energy, and to ensure that a varying equation of state does not confuse any detection of a variation in $\Geff$.  The evolution of $\fs(a)$ for each Case is shown in Fig.~\ref{fig1}, while in Fig.~\ref{fig2} we show the difference in $\fs(a)$ between each Case and the \lcdm model for $\omms=0.27$ and $\sigma_8=0.8$. Figs.~\ref{fig1}a and Figs.~\ref{fig2}a correspond to Case 1, while Figs.~\ref{fig1}b and Figs.~\ref{fig2}b correspond to Case 2. As it can be seen in Fig.~\ref{fig2}b (black solid line), the effect of an evolving equation of state is more prominent, ie the departure from the expected \lcdm model is larger, at intermediate values of the scale factor $a$, or equivalently at $z\simeq1$. Furthermore, as is clearly seen in Fig.~\ref{fig2}, a value for $g_2$ which is either negative or positive (dashed and dotted lines) causes $\fs(a)$ to depart significantly at late times from its expected \lcdm value. That is why we are able to use WiggleZ measurements of $\fs(a)$ to detect non-zero values of $g_2$ and thus $\ddotGeff$.

Having established some simulated (noise-free) data, we then attempted to recover the values of $\omms$ and $\Geff$ in that data by fitting four different models:
\begin{enumerate}
\item Model 1 ($M_1$): The analytical solution of Eq.~(\ref{Da1}) with free parameters $(\omms,w)$.  This model assumes GR is valid, and $\Geff=1$.
\item Model 2 ($M_2$): The generic approximation of Eq.~(\ref{fg}) with free parameters $(\omms,w)$ but $\gamma$ fixed to the approximation that is valid for constant equation of state $\gamma=\frac{3 (1-w)}{5-6 w}$.
\item Model 3 ($M_3$): The generic approximation of Eq.~(\ref{fg}) with free parameters $(\omms,\gamma,w)$.  This is the most flexible model and is appropriate for models with varying $w(a)$ and $\Geff(a)$.
\item Model 4 ($M_4$): A series expansion $\fs(a)=c_1 a+c_2 a^2$. This will serve as a theory-independent test of the method.
\end{enumerate}

After fitting the four different models to the data by $\chi^2$-squared minimization, we used Eqs.~(\ref{om}) and (\ref{g2}) to determine the values of $\ommgr$ and $g_2$ respectively, which we show in Table \ref{table1}. The numbers in brackets under each model show the  recovered values for $(\ommgr,g_2)$ while the first column labeled ``Original $g_2$" shows the assumed value of $g_2$ by which the data were constructed.

As it can be seen in Table \ref{table1}, $M_3$ is in all cases the most accurate in predicting the correct values for $g_2$ even in the case of an evolving equation of state $w(z)$. The magnitude of systematic error of $g_2$ ranges from $8\%$ in Case 1 (for $g_2=-1$) and less than $4\%$ for all other cases and choices of $g_2$. The reason for this is that $M_3$ is a generic approximation to the growth factor and has enough flexibility to fit a variety of different behaviors, not only an evolving equation of state such as $w(z)$ but also the more complicated Case 2 that has an evolving Newton's constant ($g_2=(-1,1)$). Surprisingly, $M_1$, which is the analytical solution of the growth factor for a universe with a constant dark energy equation of state $w(a)=w_0$, and its approximation $M_2$, do quite well in most cases with their magnitude of systematic error ranging from $14\%$ in Case 1 with $g_2=-1$ and less than $2\%$ in Case 2 with $g_2=-1$. The polynomial $M_4$ fares moderately in all cases with a magnitude of systematic error better than $40\%$ in both cases when $g_2=1$ but manages to determine $g_2$ to within $9\%$ of its true value in Case 2 (for $g_2=-1$). Finally, it is worth mentioning that the reason
the best-fitting values in Table \ref{table1} differ from the input values is that
the models being fitted are only an approximation to the growth
histories in each case, or in other words, the deviations are measuring
the level of systematic error inherent in using each model. Also, some of the error in the recovered values of $g_2$ and $\ommgr$ comes from the degeneracy between the effect of evolution of Newton's constant $\Geff$ and the effects of an evolving dark energy.

The second test of our method was to create synthetic data with noise. More specifically, we created 420 mock WiggleZ data sets $(z_i,\fs_i,\sigma_i)$, where $i=1,..,4$. For each of the four data points in the 420 synthetic data sets we used the redshift $z_i$ and error $\sigma_i$ of the real WiggleZ data, but $\fs_i$ was created by assuming a model with an evolving dark energy equation of state $w(a)=w_0+w_a (1-a)$ and a value for $g_2$ (just like in Case 2 of the noise-free test), then solving Eq.~(\ref{ode}) numerically to find $\fs_{th}(a)$ and finally adding noise, ie  $\fs_i=\fs_{th}(z_i)+N_i$. We assumed that $N_i$ is a random number drawn from a normal distribution with zero mean and a standard deviation equal to $\sigma_i$. As before, we again used the same models $(M_1, M_2, M_3, M_4)$ and considered the following cases that correspond to models with $(w_0,w_a)=(-0.95,0.5)$, $\omms=0.27$, $\sigma_8=0.8$ and three values of $g_2=(-1,0,1)$. After fitting the four different models to each of the 420 data sets implementing a $\chi^2$-statistic, we used Eqs.~(\ref{omGR}) and (\ref{g2}) to determine the values of $\ommgr$ and $g_2$ for each data set respectively. In Table \ref{table2} we present the recovered average values for the parameters $\langle \ommgr \rangle$, $\langle g_2\rangle$, $\langle \O\rangle$ and their respective $1\sigma$ errors. The $1\sigma$ errors quoted in Table \ref{table2} correspond to one standard deviation of the parameters over the 420 mock data sets.

\begin{table*}
\begin{center}
\caption{The results of the simulations with noisy data. All of the cases correspond to simulated data with $(w_0,w_a)=(-0.95,0.5)$, $\omms=0.27$, and $\sigma_8=0.8$. The columns under each model show the recovered average values for the parameters $\langle \ommgr \rangle$, $\langle g_2\rangle$, $\langle \O\rangle$ and their respective $1\sigma$ errors, while the first column labeled ``Original $g_2$" shows the assumed value of $g_2$ by which the synthetic data were constructed. \label{table2} }
\hspace{-1cm}\begin{tabular}{cccccc}
  \hline \vspace{-5pt}\\
   \vspace{5pt}Original $g_2$  &~~ Parameters ~~&$M_1$ & $M_2$ & $M_3$ & $M_4$ \\  \hline \hline \\

  \vspace{6pt}  & $\langle \ommgr \rangle$ & $0.382\pm0.097$ & $0.383\pm0.097$ & $0.388\pm0.104$ & $0.342\pm0.105$\\
  \vspace{6pt}  $g_2=-1$  & $\langle g_2\rangle$ & $-0.949\pm0.823$ & $-0.960\pm0.819$ & $-1.000\pm0.879$ & $-0.609\pm0.902$\\
  \vspace{6pt}  & $\langle \O\rangle$ &$-0.414\pm0.360$ & $-0.419\pm 0.359$ & $-0.437\pm0.385$ & $-0.266\pm0.391$\\

  \hline \\

  \vspace{6pt} & $\langle \ommgr \rangle$ & $0.294\pm0.074$ & $0.296\pm0.073$ & $0.292\pm0.076$ & $0.236\pm0.097$\\
  \vspace{6pt} $g_2=0$   & $\langle g_2\rangle$& $-0.206\pm0.640$ & $-0.218\pm0.636$ & $-0.183\pm0.663$ & $-0.304\pm0.845$\\
  \vspace{6pt}  & $\langle \O\rangle$& $-0.090 \pm 0.273$ & $-0.095\pm 0.271$ & $-0.080\pm 0.283$ & $0.127\pm 0.360$

  \\\hline \\

  \vspace{6pt} & $\langle \ommgr \rangle$& $0.180\pm0.050$ & $0.181\pm0.050$ & $0.173\pm0.053$ & $0.122\pm0.074$\\
  \vspace{6pt} $g_2=+1$   & $\langle g_2\rangle$ &$0.82\pm0.459$ & $0.805\pm0.482$ & $0.884\pm0.489$ & $1.316\pm0.717$\\
  \vspace{6pt}  & $\langle \O\rangle$ &$0.334\pm 0.185$ & $0.331\pm 0.187$ & $0.360\pm $0.197 & $0.546\pm 0.275$\\
  \hline
\end{tabular}
\end{center}
\end{table*}

All four models were able to recover the correct value for $g_2$ within $1\sigma$ of the actual value used to create the simulated data. The most accurate model, ie the one that was closer to the ``original" value of $g_2$ used to create the synthetic data, was $M_3$ which gave an exact value when $g_2=-1$ and was roughly within $10\%$ and $20\%$ when $g_2=0$ and $g_2=1$ respectively. The most precise model, ie the one with the smaller error bars, was found to be $M_2$ for $g_2=0$ and $g_2=-1$, but it was $M_1$ for $g_2=1$. Surprisingly and despite its simplicity compared to the other models, the polynomial $M_4$ proves to be equally precise when $g_2=-1$, but its precision drops significantly in the other two cases ($g_2=0$ and $g_2=1$) as the predicted error bars are larger than that of the other models by a factor of roughly 1.3 and 1.5 respectively. Also, the magnitude of the systematic error of $g_2$ is in all cases roughly within $40\%$ and $30\%$ in the first and the last two cases respectively. Therefore, we find that both the simulations with synthetic noise-free and noisy data lead to the conclusion that the model $M_3$ provides the most accurate measurements of the value of $g_2$ while having a precision comparably good to the other models. At this point  it is probably worth noting that the average values of $\ommgr$
and $g_2$ in Tables \ref{table1} and \ref{table2}  agree reasonably well each other  for each case respectively.

Finally, as it can be seen in both Tables \ref{table1} and \ref{table2}, this method seems to be overestimating the value for $\ommgr$ when $g_2<0$ and underestimates it when $g_2>0$. So, when the value of $\omms$ measured by independent techniques disagrees with the value of $\ommgr$ recovered using our technique, then this can be interpreted as evidence for a varying Newton's constant. This is in agreement with the expected behavior of the ratio $r$ of Eq.~(\ref{ratiog2}). Furthermore, in all cases the sign of the average value of $\O$ over the models was consistent with the assumed value of $g_2$. The $\O$ diagnostic would allow for a statistically-significant detection of varying $\Geff$ when $g_2 = \pm 1$ given higher signal-to-noise growth measurements.

\subsection{WiggleZ data analysis}

\begin{table*}[t!]
\begin{center}
\caption{The results of the analysis of the WiggleZ data for $g_2$, $\ommgr$ and the diagnostic $\O$. When $\O \neq 0$, then this indicates an evolving $\Geff$. The evolution of $\fs(a)$ for the best-fit models $(M_1, M_2, M_3, M_4)$ and their $1\sigma$ error region (gray shaded area) along with a \lcdm model for $\omms=0.27$ (green dashed line) can be seen in Fig.~\ref{fig3}. In each case the first row is the estimated result and the error using the Fisher analysis, while the second and the third rows correspond to the error estimated by simulated data-sets and the jack-knife respectively (see Appendix \ref{errap} for more details). The discrepancy of the jack-knife errors and the ones estimated by the other two approaches is explained in the text. \label{table3}}\vspace{5pt}
\begin{tabular}{cccccc}
 \hline \vspace{-5pt}\\
  \vspace{5pt} & $M_1$  & $M_2$ & $M_3$ & $M_{3, w=-1}$ & $M_4$ \\
  \hline\hline \\
  $\Omega_{\rm m,GR}$ & $~0.399^{+0.095}_{-0.106}$ & $~0.400^{+0.094}_{-0.105}$ & $~0.389^{+0.126}_{-0.136}$ & $~0.417^{+0.083}_{-0.094}$& $~0.379^{+0.111}_{-0.128}$\vspace{5pt}\\
  Simulations& $\pm0.095$ & $\pm0.091$ & $\pm0.134$ & $\pm0.079$& $\pm0.120$\vspace{5pt}\\
  Jack-knife& $\pm0.183$ & $\pm0.182$ & $\pm0.195$ & $\pm0.124$& $\pm0.246$\vspace{5pt}\\
  \hline \\
  $g_2$ & $-1.120^{+0.920}_{-0.822}$ & $-1.128^{+0.909}_{-0.810}$ & $-1.029^{+1.177}_{-1.085}$ & $-1.268^{+0.813}_{-0.715}$& $-0.939^{+1.103}_{-0.961}$\vspace{5pt}\\
  Simulations & $\pm0.890$& $\pm0.923$ & $\pm1.217$ & $\pm0.821$& $\pm1.215$\vspace{5pt}\\
  Jack-knife & $\pm1.554$ & $\pm1.546$ & $\pm1.655$ & $\pm1.063$& $\pm2.076$\vspace{5pt}\\
  \hline \\
  $\O$ & $-0.48^{+0.35}_{-0.39}$ & $-0.48^{+0.35}_{-0.39}$ & $-0.44^{+0.47}_{-0.50}$ & $-0.54^{+0.31}_{-0.35}$& $-0.40^{+0.41}_{-0.47}$\vspace{5pt}\\
  Simulations & $\pm0.354$ & $\pm0.349$ & $\pm0.550$ & $\pm0.328$& $\pm0.410$\vspace{5pt}\\
  Jack-knife & $\pm0.678$ & $\pm0.675$ & $\pm0.720$ & $\pm0.460$&$\pm0.913$\vspace{5pt}\\

  \hline\hline \\

\end{tabular}
\end{center}
\end{table*}

In this section we will present the results of the method we described in the previous sections, when it is applied to the real WiggleZ data. After fitting the four models $(M_1, M_2, M_3, M_4)$ to the data, we used Eqs.~(\ref{omGR}), (\ref{g2}) and (\ref{diag}) to determine the values of $\ommgr$, $g_2$ and $\O$ respectively, which we show in Table \ref{table3}. The errors of the values of the parameters were determined by a Fisher matrix analysis, see Chapter 15 of Ref.~\cite{Press:1992zz} for more details on the method.

We had to perform a Fisher matrix analysis since we cannot estimate the errors on the parameters with the simulated datasets of the previous section. The reason for this is the specific way these synthetic data were constructed. In fact, there are two ways one can create synthetic data:

\begin{itemize}
  \item One can assume a theoretical model, eg have specific choices for $w(a)$ and $\Geff(a)$, calculate $\fs$ at the WiggleZ redshifts and add gaussian noise. We then use these mock data to test whether the proposed method works as expected, ie it recovers the original and ``correct" values. This can only be used to determine if the method works and not to derive the errors of parameters for the real data, since we did not use the latter at all. This is what we did in the previous section. Since our mock data resemble the real ones, the errors will be as well similar.
  \item The second way is to draw N data points from the real data with replacement, ie do a bootstrap Monte Carlo (see Chapters 15.6.1-15.6.2 of Ref.~\cite{Press:1992zz}). This way we get datasets in which a random fraction of the original points are replaced by duplicated original points. Then one subjects the synthesized data to the same method and derives the desired parameters for each mock set. The standard deviation of the parameters corresponds to the error of the parameters deduced from the real data.
\end{itemize}

Clearly, these two methods test different things. In our case, we cannot use the second method (the Bootstrap Monte Carlo) to determine the errors of the desired parameters as our real dataset is too small, so we had to resort to Fisher matrix analysis.

In order to confirm the Fisher matrix analysis we considered the jack-knife approach to error estimation, see Ref. \cite{efron} for an introduction, and the variance between several simulations, see Chapter 15.6 of Ref. \cite{Press:1992zz}. We give more details on how we implemented these two methods in Appendix \ref{errap}. We found that the method of calculating the variance between several simulations of mock data is in good agreement with the Fisher estimates of Table \ref{table3}. On the contrary, the estimated error-bars from the jack-knife method are significantly larger than the ones obtained by the other two methods and this is mainly for two reasons:

\begin{itemize}
  \item Our estimators Eqs. (\ref{omGR}), (\ref{omreal}) and (\ref{g2}) are not smooth statistics. For example, it is well known that the jack-knife works very well for statistics like the mean but fails for the median (see Ref. \cite{efron}).

  \item The size of the data-set is quite small, so this affects the fitting procedure (deleting one point out of four data-points results in a data-set which is barely big enough to be fitted by Model $M_3$ that has 3 free parameters.
\end{itemize}

\begin{figure*}[t!]
\vspace{0cm}\rotatebox{0}{\vspace{0cm}\hspace{-0.5cm}\resizebox{1\textwidth}{!}{\includegraphics{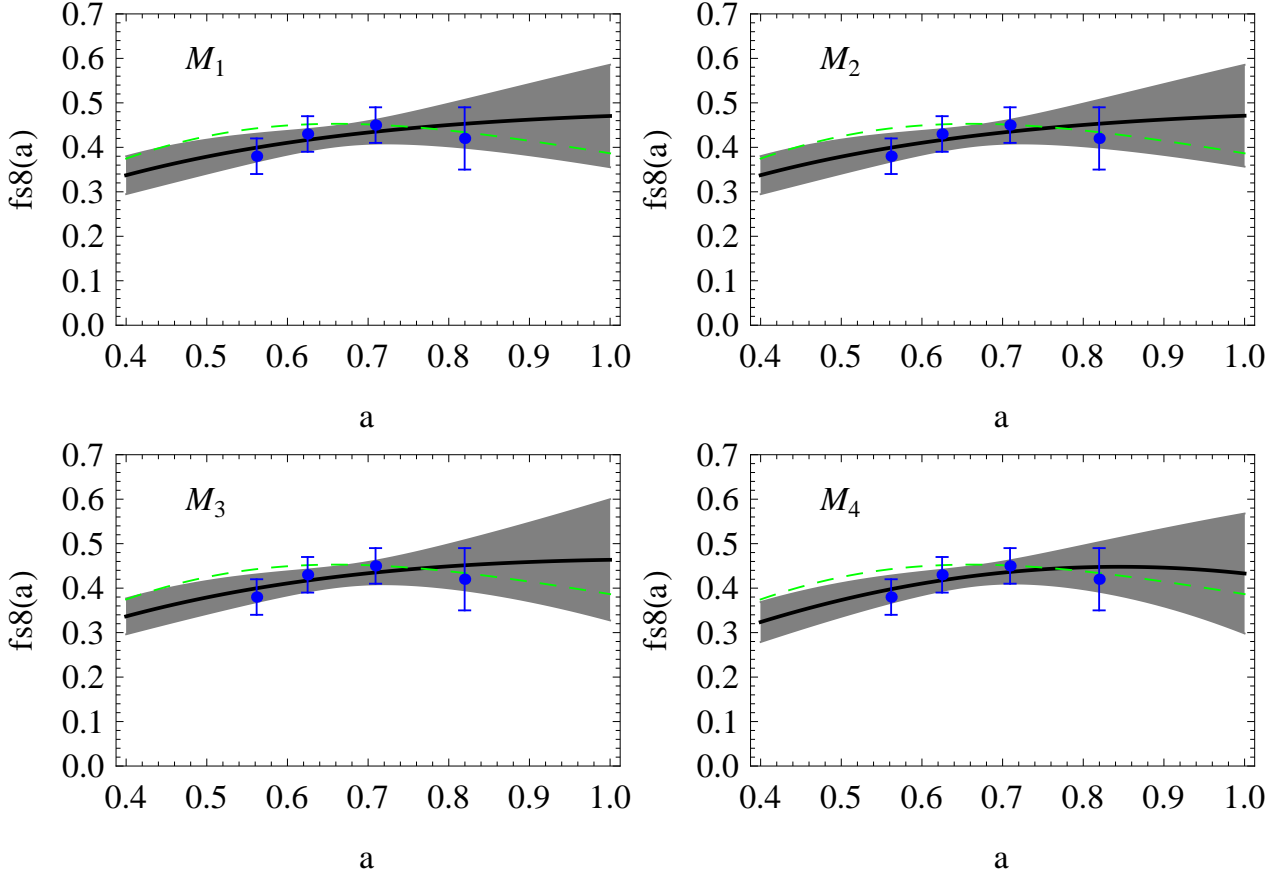}}}
\caption{The evolution of $\fs(a)$ for the best-fit models $(M_1, M_2, M_3, M_4)$ and their $1\sigma$ error region (gray shaded area) along with a \lcdm model for $\omms=0.27$ (green dashed line).\label{fig3}}
\end{figure*}

The evolution of $\fs(a)$ for the four best-fit models and their $1\sigma$ error region (gray shaded area) along with a \lcdm model for $\omms=0.27$ (green dashed line) can be seen in Fig.~\ref{fig3}. The most precise measurement of $g_2$, ie the one that has the smaller error bars, is that of the model $M_2$ that predicts $g_2=-1.128^{+0.909}_{-0.810}$, which offers no convincing evidence for non-zero $g_2$. As expected, $M_1$ predicts similar results to $M_2$, but the value of $g_2$ for $M_3$ has slightly larger errors, making its prediction even more compatible with GR.

When we fix the background cosmology to \lcdm, ie set $w=-1$, but allow the parameter $\gamma$ to be free (case $M_{3, w=-1}$ in Table \ref{table3}) in order to capture the effects of an evolving $\Geff$, then we get a prediction which again offers no convincing evidence for a non-zero $g_2$. The models $M_1$, $M_2$, $M_3$ and $M_{3, w=-1}$ give on average a value $g_2\simeq-1.14 \pm 0.91$ which corresponds to a value for the second derivative of $\ddotGeff(t_0)=-1.19\pm 0.95\cdot 10^{-20}h^2 \textrm{yr}^{-2}$. Also, their prediction for the matter density ($\ommgr=0.40\pm0.10$) is in agreement with the one measured from the WMAP probe ($\omms=0.27\pm0.03$) \cite{Komatsu:2010fb}. This is also in agreement with the results of the diagnostic $\O=-0.486\pm0.369$, which is compatible with zero. We should note that while the estimate of $g_2$ depends on the assumption of a slowly varying $\Geff$ on the form of Eq.~(\ref{geff}), the estimate of $\O$ is completely model independent of any assumption on either $\Geff$ or the dark energy equation of state $w(a)$.

The theory-independent model $M_4$, $\fs(a)=c_1 a+c_2 a^2$, gives a value $g_2=-0.939^{+1.103}_{-0.961}$ which is in agreement with the other models, albeit with a slightly larger upper error, which makes it consistent with GR. Also, $M_4$
can be used to get an estimate for $\sigma_8$. Remembering the definition $\fs(a)\equiv f(a) \sigma_8(a)=\frac{\sigma_8}{\delta(1)}~a~\delta'(a)$ and the fact that for $a\ll1$ we have $\delta'(a)\simeq1+O(a)$, it is easy to see that $c_1=\frac{\sigma_8}{\delta(1)}$, so by using Eq.~(\ref{growth}) we get $\delta(1)=1+\frac{c_2}{2 c_1}$  and $\sigma_8=c_1+c_2/2$. Using the best-fit values for $c_1$ and $c_2$ we get the result $\sigma_8=0.75 \pm 0.08$, which is in agreement with the corresponding WMAP 7 value $\sigma_8=0.80 \pm 0.03$ \cite{Komatsu:2010fb}.

We also repeated our calculations of $g_2$ for data that were marginalized over an Alcock-Paczynski scale distortion factor \cite{park} and found that the results degraded only by $10\%$, so our conclusions remained unchanged.

\begin{table*}[!t]
\begin{center}
\caption{The results of the joint analysis of the WiggleZ+SDSS-II LRG data for $g_2$, $\ommgr$ and the diagnostic $\O$. When $\O \neq 0$, then this means an evolving $\Geff$. The models $M_1$, $M_2$, $M_3$ and $M_{3, w=-1}$ give on average a value $g_2\simeq-0.34 \pm 0.65$. The results for $\O$ are consistent with GR at the $1\sigma$ level. In each case the first row is the estimated result and the error using the Fisher analysis, while the second and the third rows correspond to the error estimated by simulated data-sets and the jack-knife respectively (see Appendix \ref{errap} for more details). The discrepancy of the jack-knife errors and the ones estimated by the other two approaches is explained in the text. \label{table4}}\vspace{2pt}
\begin{tabular}{cccccc}
 \hline \vspace{-5pt}\\
   \vspace{5pt} & $M_1$         & $M_2$        & $M_3$ & $M_{3, w=-1}$        & $M_4$ \\
  \hline\hline \\
  $\ommgr$ & $~0.310^{+0.063}_{-0.066}$ & $~0.312^{+0.062}_{-0.065}$ & $~0.291^{+0.096}_{-0.097}$ & $~0.324^{+0.071}_{-0.073}$& $~0.261^{+0.068}_{-0.075}$\vspace{3pt}\\
  Simulations & $\pm0.069$ & $\pm0.068$ & $\pm0.09$ & $\pm0.071$& $\pm0.069$\vspace{3pt}\\
  Jack-knife & $\pm0.107$ & $\pm0.106$ & $\pm0.118$ & $\pm0.123$& $\pm0.138$\vspace{3pt}\\
  \hline \\
  $g_2$ & $-0.351^{+0.579}_{-0.544}$ & $-0.363^{+0.571}_{-0.536}$ & $-0.184^{+0.848}_{-0.836}$ & $-0.471^{+0.640}_{-0.612}$& $0.076^{+0.648}_{-0.591}$\vspace{3pt}\\
  Simulations & $\pm0.591$ & $\pm0.587$ & $\pm0.786$ & $\pm0.614$& $\pm0.581$\vspace{3pt}\\
  Jack-knife& $\pm0.928$ & $\pm0.924$ & $\pm1.027$ & $\pm1.072$& $\pm1.207$\vspace{3pt}\\
  \hline \\
  $\O$ & $-0.15^{+0.23}_{-0.24}$ & $-0.16^{+0.23}_{-0.24}$ & $-0.08^{+0.36}_{-0.36}$ & $-0.20^{+0.26}_{-0.27}$& $0.03^{+0.25}_{-0.28}$\vspace{3pt}\\
  Simulations & $\pm0.263$ & $\pm0.261$ & $\pm0.329$ & $\pm0.256$& $\pm0.259$\vspace{3pt}\\
  Jack-knife & $\pm0.395$ & $\pm0.394$ & $\pm0.436$ & $\pm0.457$& $\pm0.510$\vspace{3pt}\\
  \hline
\end{tabular}
\end{center}
\end{table*}

Finally, the error estimates for $g_2$ might be somewhat large but the big advantage of this method is the fact that we have not used any other data besides the four data points from the WiggleZ survey. Also, it is worth emphasizing the point that the
results we get are independent of the model used, in that the systematic
differences in $\ommgr$, $g_2$ and $\O$ between the models are much smaller
than the statistical errors in these measurements. Therefore, the only thing that limits the accuracy of this method is the precision of the data and the number of redshift bins. We will explore this issue in what follows.

\section{Other datasets and forecasts}

\begin{figure*}[!t]
\vspace{0cm}\rotatebox{0}{\vspace{0cm}\resizebox{.47\textwidth}{!}{\includegraphics{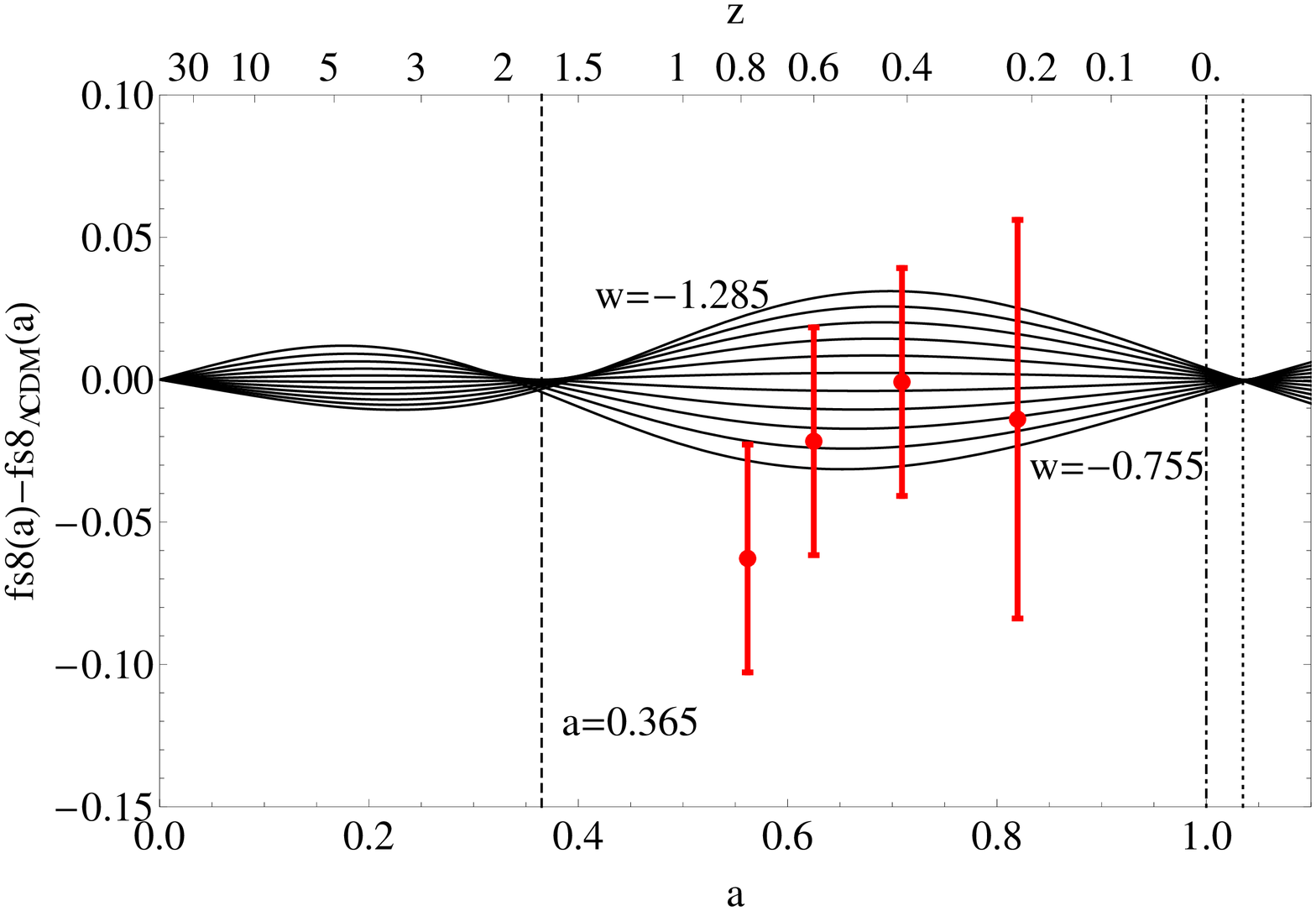}}}
\hspace{0.25cm}
\vspace{0cm}\rotatebox{0}{\vspace{0cm}\resizebox{.47\textwidth}{!}{\includegraphics{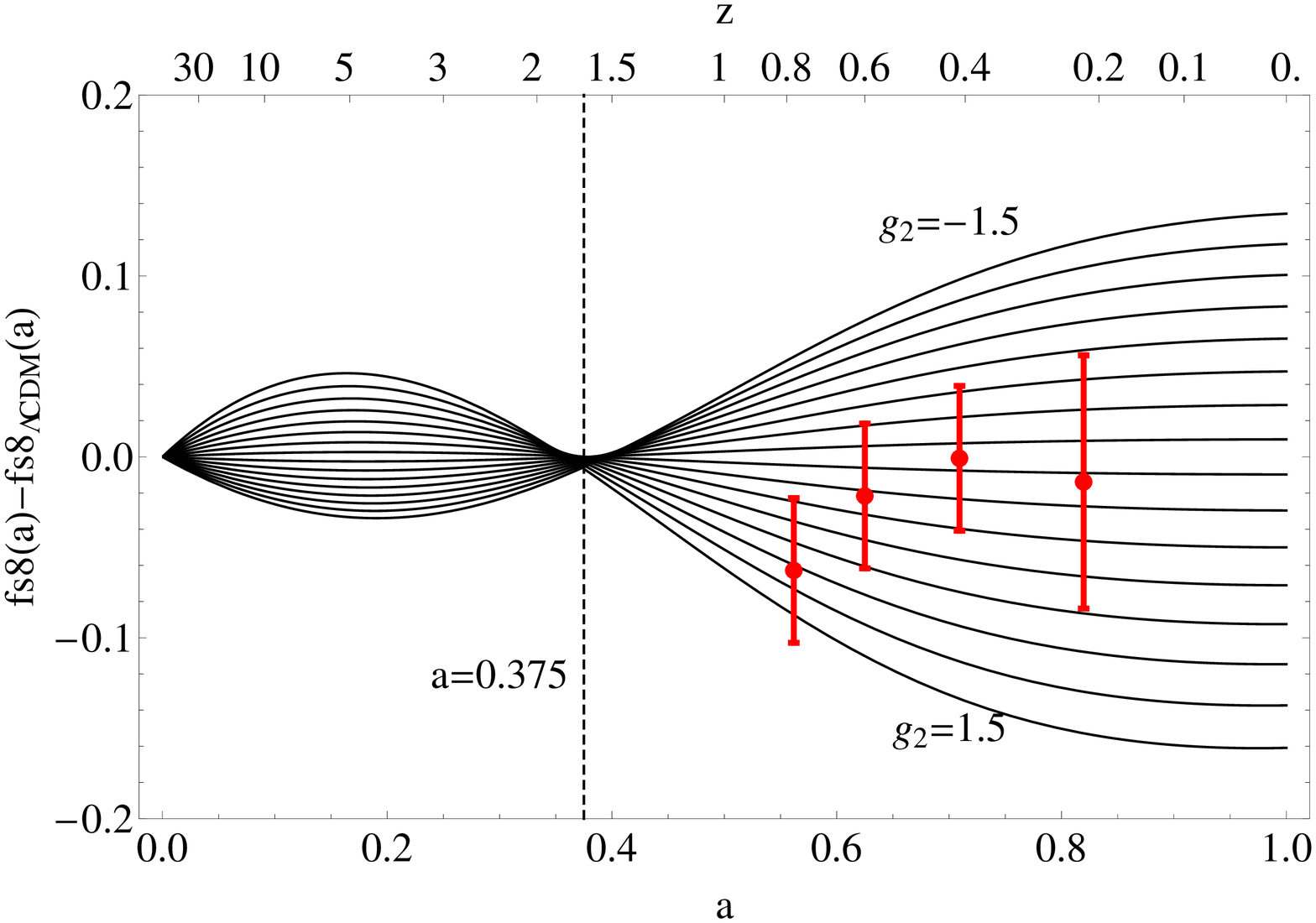}}}
\caption{Left: The difference in the evolution of $\fs(a)$, for values of $w$ in the range $w\in[-1.285,-0.755]$ in steps of $\delta w=0.053$ (from top to bottom), and the \lcdm model. The range  of $w$ corresponds to five $\sigma$'s from the WMAP7+BAO+SN Ia best fit $w=-1.02\pm0.053$ of Ref.~\cite{Komatsu:2010fb}. Right: The difference in the evolution of $\fs(a)$ with $w(z)=-1$ and an evolving $\Geff$ for values of $g_2$ in the range $g_2\in[-1.5,1.5]$ and in steps of $\delta g_2=0.2$ (from top to bottom) and the \lcdm model. Both plots were created with $\omms=0.27$ and $\sigma_8=0.8$, while the red points correspond to the real WiggleZ data. In the plot on the left, there are two sweet-spots, one at $a=0.365$ or $z=1.74$ (dashed vertical line) and another at $a=1.035$ or $z=-0.034$ (dotted vertical line) which takes place in the future (the present time $a=1$ corresponds to the dashed-dotted line). On the plot on the right there is only one sweet-spot at $a=0.375$ or $z=1.67$ (dashed vertical line). Therefore, having more points at low redshifts ($z<0.8$) will help in the detection of an evolving $\Geff$ and its discrimination from $w\neq-1$. In these plots the redshift $z=0.8$ corresponds to $a\simeq0.56$, which is near the data point that is furthest to the left. \label{fig4}}
\end{figure*}

In this section we present the results from a joint analysis of the growth rate measurements from the WiggleZ and the SDSS-II data of Ref.~\cite{Samushia:2011cs}, along with the optimal redshift range where new data points should be attempted to be measured in future surveys in order to achieve the lowest error in $g_2$.

By constraining the large-scale redshift-space distortions of the LRG SDSS-II data, Ref.~\cite{Samushia:2011cs} presented two new data points  $f(z=0.25)\sigma_8(z=0.25)=0.3930 \pm 0.0457$ and $f(z=0.37)\sigma_8(z=0.37)=0.4328 \pm 0.0370$. After fitting the four models $(M_1, M_2, M_3, M_4)$ to the joint WiggleZ and LRG SDSS-II data, we used Eqs.~(\ref{omGR}), (\ref{g2}) and (\ref{diag})  to determine the values of $\omms$, $g_2$ and $\O$ respectively, which we show in Table \ref{table4}. The errors of the parameters were determined by a Fisher matrix analysis, see Chapter 15 of Ref.~\cite{Press:1992zz} for more details on the method.

The most precise measurement of $g_2$, ie the one that has the smaller error bars, is that of the model $M_2$ that predicts $g_2=-0.363^{+0.571}_{-0.536}$, which is compatible with $g_2=0$ at the $1\sigma$ level. As expected, $M_1$ predicts similar results to $M_2$, but the value of $g_2$ for $M_3$ has much larger errors, thus making its prediction compatible with both GR and $g_2=-1$ at the $1\sigma$ level. When we fix the background cosmology to \lcdm, ie set $w=-1$, but allow the parameter $\gamma$ to be free (case $M_{3, w=-1}$ in Table \ref{table4}) in order to capture the effects of an evolving $\Geff$, then we get a prediction for $g_2$ which is compatible with both GR and $g_2=-1$ at the $1\sigma$ level. The models $M_1$, $M_2$, $M_3$ and $M_{3, w=-1}$ give on average a value $g_2\simeq-0.34 \pm 0.65$ which corresponds to a value for the second derivative of $\ddotGeff(t_0)=-3.6\pm 6.8\cdot 10^{-21}h^2 \textrm{yr}^{-2}$ and is again compatible with GR. Also, in this case all models are compatible with $\O=0$ at the $1\sigma$ level. Finally, the polynomial $M_4$ gives a measurement of $\sigma_8=0.77 \pm 0.07$, which is in agreement with the corresponding WMAP 7 value $\sigma_8=0.80 \pm 0.03$ \cite{Komatsu:2010fb}.

However, we should note that one should be careful when combining two different data sets as this may result in systematic errors. For example, this was the case with the ``Gold06'' set, which was one of the early SN Ia data sets and was made of several different subsets \cite{Nesseris:2006ey}. For the case at hand, one reason to exercise caution would be some low level of correlation due to observing some common areas of sky or the different redshift-space distortion models used by the two teams. Another systematic issue might be the galaxy bias, since the two surveys use two different types of tracers (star-forming galaxies and LRGs for the WiggleZ and SDSS respectively), but both teams have explored this assumption in detail \cite{Blake:2011rj, Samushia:2011cs}.

We also tried to determine the optimal redshift range where new data points should be attempted to be measured in future surveys. We noticed that due to degeneracies in $\fs(a)$ there are several sweet-spots at specific redshifts, ie there are points where $\fs(a)$ is the same regardless of the values of the parameters $w$ or $g_2$. In Fig.~(\ref{fig4}) (left) we show the difference in the evolution of $\fs(a)$, for values of $w$ in the range $w\in[-1.285,-0.755]$ in steps of $\delta w=0.053$ (from top to bottom), and the \lcdm model. The range  of $w$ corresponds to five $\sigma$'s from the WMAP7+BAO+SN Ia best fit $w=-1.02\pm0.053$ of Ref.~\cite{Komatsu:2010fb}. In Fig.~\ref{fig4} (right) we show the difference in the evolution of $\fs(a)$ with $w(z)=-1$ and an evolving $\Geff$ for values of $g_2$ in the range $g_2\in[-1.5,1.5]$ and in steps of $\delta g_2=0.2$ (from top to bottom) and the \lcdm model. Both plots were created with $\omms=0.27$ and $\sigma_8=0.8$, while the red points correspond to the real WiggleZ data. In the plot on the left, there are two sweet-spots, one at $a=0.365$ or $z=1.74$ (dashed vertical line) and another at $a=1.035$ or $z=-0.034$ (dotted vertical line) which takes place in the future (the present time $a=1$ corresponds to the dashed-dotted line). On the plot on the right there is only one sweet-spot at $a=0.375$ or $z=1.67$ (dashed vertical line). These sweet spots are really sour spots, as observations in these regions cannot distinguish between different values of $w$ and $g_2$. It is therefore clear that having more points at low redshifts ($0<z<0.8$) will help in the detection of an evolving $\Geff$ and its discrimination from $w\neq-1$. In these plots the redshift $z=0.8$ corresponds to $a\simeq0.56$, which is near the data point that is furthest to the left.

\begin{figure*}[!t]
\vspace{0cm}\rotatebox{0}{\vspace{0cm}\resizebox{.47\textwidth}{!}{\includegraphics{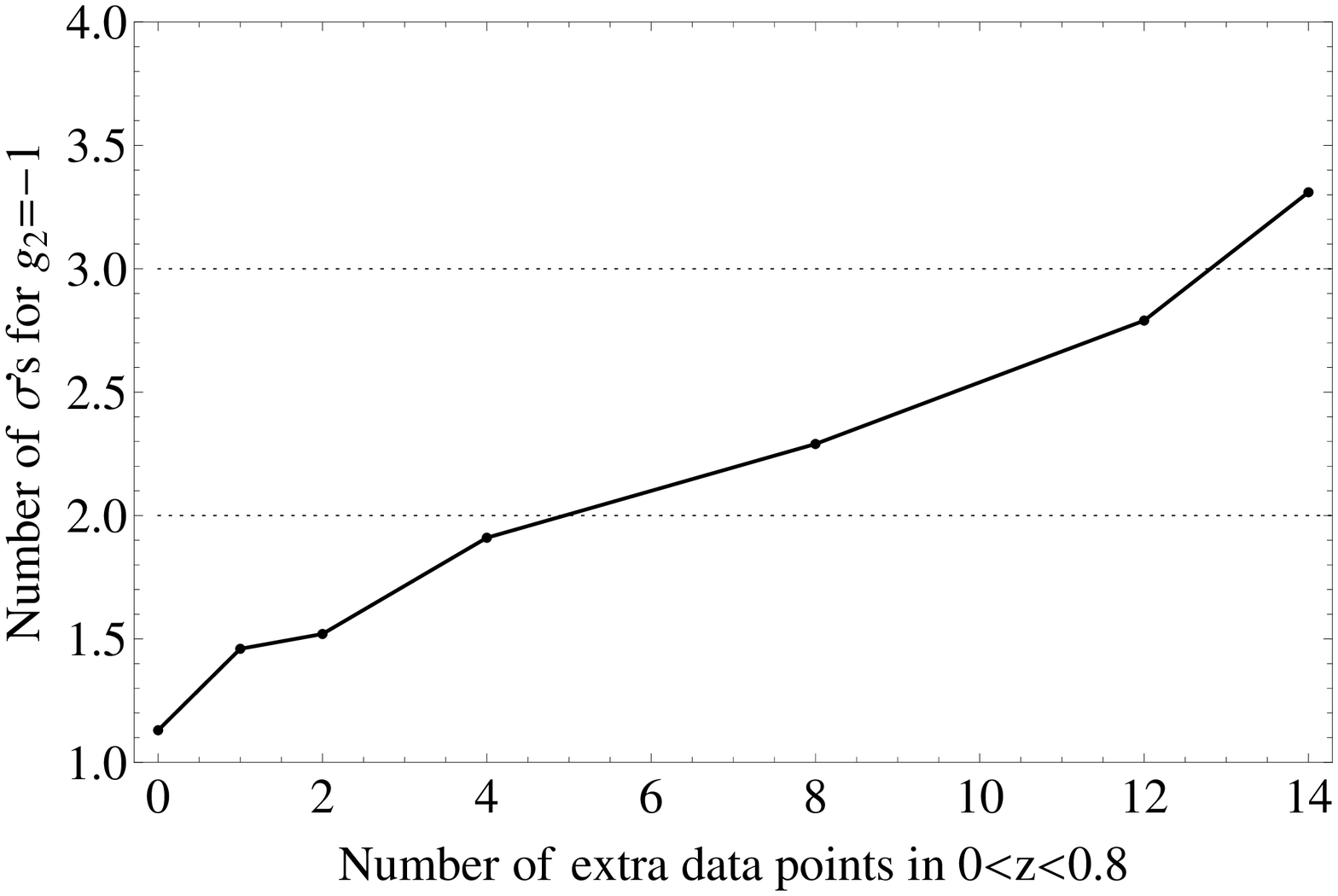}}}
\hspace{0.25cm}
\vspace{0cm}\rotatebox{0}{\vspace{0cm}\resizebox{.47\textwidth}{!}{\includegraphics{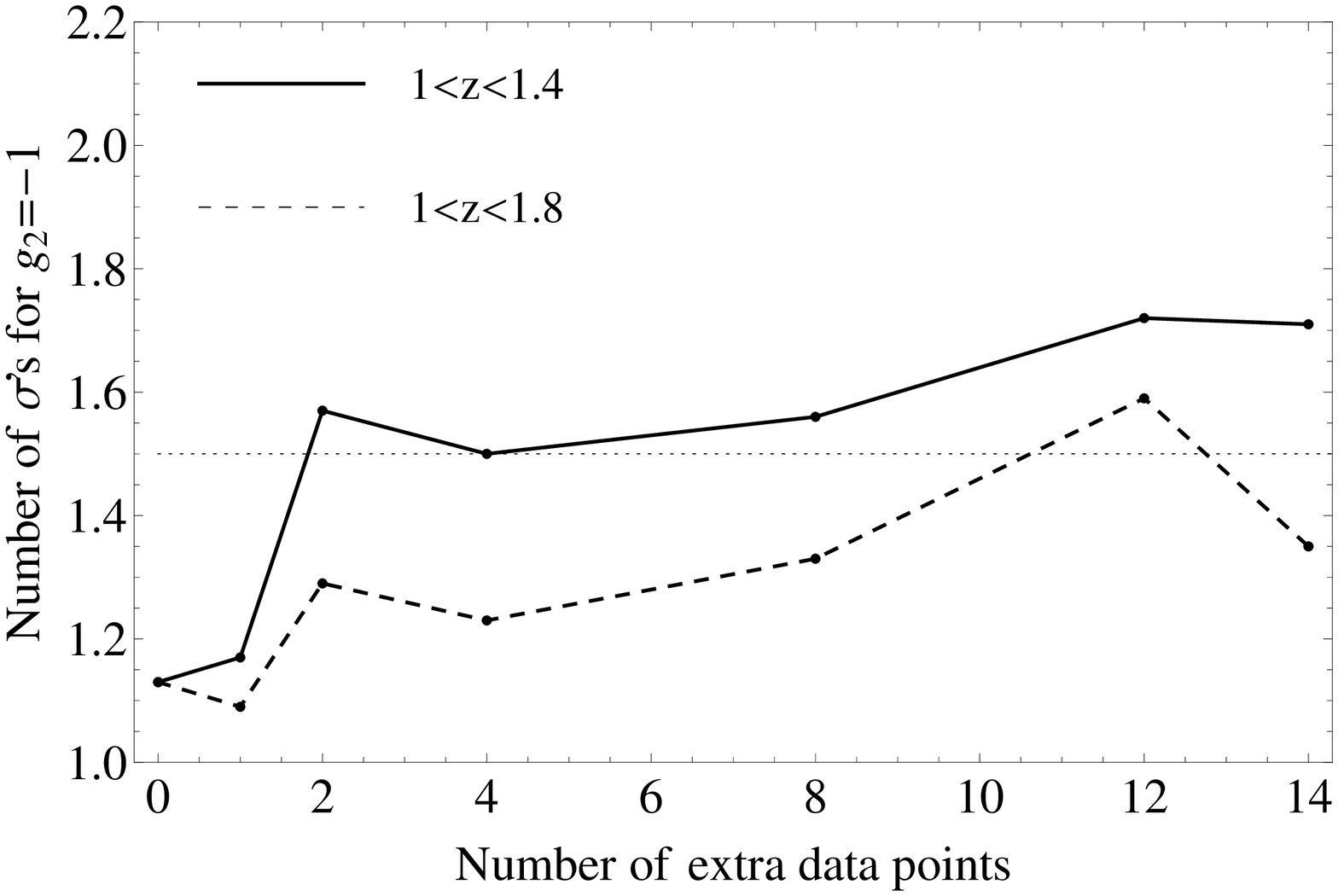}}}
\caption{Left: The number of extra data points in the range $0<z<0.8$ vs the corresponding number of $\sigma$'s in a detection of $g_2=-1$. For a $2\sigma$ detection of $g_2=-1$, 5 new points are needed in the range $0<z<0.8$, while for a $3\sigma$ detection 13 new points are needed. Right: The number of extra data points in two ranges $1<z<1.4$ (solid line) and $1<z<1.8$ (dashed line) vs the corresponding number of $\sigma$'s in a detection of $g_2=-1$. The data for both plots were created with the same parameters as the synthetic data of the first row of Table \ref{table2} (the data with $g_2=-1$). Having more points in the redshift range $1<z<1.8$  may give at most a $1.6\sigma$ detection of $g_2=-1$, while having more points in the redshift range $1<z<1.4$ may give a detection up to $1.7\sigma$'s. The reason for this is the existence of the sour spot at $z\simeq1.7$ described in Fig.~\ref{fig4}. \label{fig5}}
\end{figure*}

In order to test whether having more points at low redshifts can help in the detection of an evolving $\Geff$ and to estimate the necessary number of new points necessary to achieve this, we created synthetic data with noise following the same procedure as in Section IV. We placed 4 of the points fixed at the WiggleZ redshifts $z=(0.22, 0.41, 0.60, 0.78)$ and we also added 1, 2, 4, 8, 12 or 14 points randomly placed in three redshift ranges
\begin{itemize}
  \item Range 1: $0<z<0.8$
  \item Range 2: $1<z<1.4$
  \item Range 3: $1<z<1.8$
\end{itemize} In order to keep the analysis as simple as possible, each of the new points was assigned an error similar to that of the real data at the corresponding redshifts, ie $\sigma=0.04$ for $z>0.3$ and $\sigma=0.07$ around $z\leq0.3$. In total we created 18 groups of 96 synthetic data sets each, all with initial conditions $g_2=-1$, $\omms=0.27$ and the rest of the parameters the same as in Section IV. We then used Eq.~(\ref{g2}) to determine the values of $g_2$ and calculate the number of $\sigma$'s to which a non-zero $g_2$ is detected. In Fig.~(\ref{fig5}) (left) we show the number of extra data points in the range $0<z<0.8$ vs the corresponding number of $\sigma$'s in a detection of $g_2=-1$. For a $2\sigma$ detection of $g_2=-1$, 5 new points are needed in the range $0<z<0.8$, while for a $3\sigma$ detection 13 new points are needed. Future observations in the redshift range $0 < z < 0.8$, besides the WiggleZ and the LRG SDSS data, also include the Baryon Oscillation Spectroscopic Survey (BOSS). Specifically, they study the correlation functions of massive galaxies in the redshift range $0<z<0.7$ and aim to make a precise determination of the scale of baryon oscillations and to study the growth of structure and the evolution of massive galaxies \cite{White:2010ed}.

In Fig.~\ref{fig5} (right) we show the number of extra data points in two ranges $1<z<1.4$ (solid line) and $1<z<1.8$ (dashed line) vs the corresponding number of $\sigma$'s needed for a detection of $g_2=-1$. The data for both plots were created with the same parameters as the synthetic data of the first row of Table \ref{table2} (the data with $g_2=-1$). Having more points in the redshift range $1<z<1.8$  may give at maximum an $1.6\sigma$ detection of $g_2=-1$, while having more points in the redshift range $1<z<1.4$ may give a detection up to $1.7\sigma$'s. The reason for this is the existence of the sour spot at $z\simeq1.7$ described in Fig.~\ref{fig4}. It is therefore obvious that we need both more and better data points in order to possibly detect an evolving $\Geff$ with any statistical significance and that low-redshift ($z<0.8$) data is most valuable.

\section{Conclusions}
We presented a method that can be used to determine the value of the second derivative of Newton's constant $\ddotGeff(t_0)$, under the assumption of a slowly varying $\Geff(a)$, by using the growth rate data of the WiggleZ survey. The novelty of our approach lies in the fact that contrary to other methods, this one does not require knowledge of the expansion history of the Universe $H(z)$, usually found separately by using the Type Ia Supernovae data but instead we can use the generic approximation of Eq.~(\ref{fg}) which, as we have shown in the previous section, is able to fit a great variety of different scenarios (evolving equation of state and/or an evolving Newton's constant).

In order to demonstrate the ability of our method to measure the correct values of $g_2$ we performed simulations with synthetic noise-free and noisy data. We used four ``fiducial" models, the analytic solution to Eq.~(\ref{ode}) $M_1$, the generic approximation of Eq.~(\ref{fg}) with the parameter $\gamma$ fixed $M_2$ or free to vary $M_3$ and a polynomial expansion $M_4$. We tested the models with data created for different parameters $(w_0,w_a)$ of a dark energy equation of state $w(a)=w_0+w_a(1-a)$, three different values of the second derivative of Newton's constant $g_2 \equiv \ddotGeff(t_0)/(\Geff(t_0)H_0^2)=(-1,0,1)$.

All models proved to be able to measure the correct value of $g_2$ within $1\sigma$ of its true value, despite the fact that we tested them against data created from an evolving equation of state $w(z)$. The most accurate model, ie the one that was closer to the ``true" value of $g_2$ used to create the synthetic data, was $M_3$ which gave an exact value when $g_2=-1$ and was roughly within $10\%$ and $20\%$ in the $g_2=0$ and $g_2=1$ respectively. The most precise model, ie the one with the smaller error bars, was found to be $M_2$ for $g_2=0$ and $g_2=-1$, but $M_1$ for $g_2=1$. So, we found that both the simulations with synthetic noise-free and noisy data lead to the conclusion that the model $M_3$ provides the most accurate measurements of the value of $g_2$ while having a precision comparably good to the other models.

When we applied our method to the real WiggleZ data, we found that the models $M_1$, $M_2$, $M_3$ and $M_{3, w=-1}$ give on average a value $g_2\simeq-1.14 \pm 0.91$ which corresponds to a value for the second derivative of $\ddotGeff(t_0)=-1.19\pm 0.95\cdot 10^{-20}h^2 \textrm{yr}^{-2}$. Also, their prediction for the matter density ($\ommgr=0.40\pm0.10$) is in agreement with the one measured from the WMAP probe ($\omms=0.27\pm0.03$) \cite{Komatsu:2010fb}. Finally, we found that the polynomial $M_4$ gives a measurement of $\sigma_8=0.75 \pm 0.08$, which is in agreement with the corresponding WMAP 7 value $\sigma_8=0.80 \pm 0.03$ \cite{Komatsu:2010fb}.

When we performed a joint analysis of the WiggleZ and the SDSS-II data of Ref.~\cite{Samushia:2011cs}, the models $M_1$, $M_2$, $M_3$ and $M_{3, w=-1}$ gave on average a value $g_2\simeq-0.34 \pm 0.65$ which corresponds to a value for the second derivative of $\ddotGeff(t_0)=-3.6\pm 6.8\cdot 10^{-21}h^2 \textrm{yr}^{-2}$ and is again compatible with GR. Finally, the polynomial $M_4$ gives a measurement of $\sigma_8=0.77 \pm 0.07$, which is in agreement with the corresponding WMAP 7 value $\sigma_8=0.80 \pm 0.03$ \cite{Komatsu:2010fb}.

We also presented a diagnostic \be {\O}\equiv \frac{\omms-\Omega_{\rm m,GR}}{\omms}=1-\frac{\Omega_{\rm m,GR}}{\omms}=1-\frac{1}{3 I_0 \omms} \nn\ee where $\omms$ is the value of the matter density as measured from other independent observations. As is easily seen the value of $\O$ is zero only in GR and non-zero in general in the presence of an evolving $\Geff$. Furthermore, we found that the sign of $\O$ is correlated to the sign of the second derivative $\ddotGeff$, ie they are both either positive or negative. Therefore, if we assume that the value of $\omms$ is independently and accurately determined by other observations, then any statistically significant deviation of the quantity ${\O}$ from zero, clearly and uniquely identifies the presence of an evolving $\Geff$ and consequently modified gravity theories. We should note that while the estimate of $g_2$ depends on the assumption of a slowly varying $\Geff$ on the form of Eq.~(\ref{geff}), the estimate of $\O$ is model independent of any assumption on either  $\Geff$ or the dark energy equation of state $w(a)$. Applying it to the real WiggleZ data we found $\O=-0.486\pm0.369$, which is consistent with zero.

Finally, we also determined the optimal redshift range where new data points should be attempted to be measured in future surveys. We noticed that due to degeneracies in $\fs(a)$ there are sour-spots at high redshifts, ie there points where $\fs(a)$ is the same regardless of the values of the parameters $w$ or $g_2$. So, in terms of adding new data points, we found that these should be focused in the range $0 < z < 0.8$.

In order to test whether having more points at low redshifts can help in the detection of an evolving $\Geff$ and to estimate the necessary number of new points necessary to achieve this, we created synthetic data with noise following the same procedure as in Section IV. We placed 4 of the points fixed at the WiggleZ redshifts $z=(0.22, 0.41, 0.60, 0.78)$ and we also added 1, 2, 4, 8, 12 or 14 points randomly placed in three redshift ranges $0<z<0.8$, $1<z<1.4$ and $1<z<1.8$. Having more data points in any of the tested redshift ranges of $z>1$ allowed a maximum 1.7$\sigma$ detection of $g_2=-1$, while more sticking to the lower redshift range of $0<z<0.8$ allowed a much more significant detection, at 3$\sigma$, of $g_2=-1$.

In conclusion, we presented a method by which growth of structure could detect a variation in $\Geff$ in a model-independent manner, but found that the current growth rate data is consistent with a non-evolving value of $\Geff$. It is therefore imperative to have both more and better data points in order to possibly detect an evolving $\Geff$ with any statistical significance. This makes it tantalizing to pursue this analysis further as both the quality and the number of the growth rate data will increase in the near future.

\section*{Acknowledgements}
We acknowledge financial support from the Australian Research Council
through a Discovery Project grant funding the position of DP. SN acknowledges support by the Niels Bohr International Academy, the Danish Research Council (FNU) under FNU Grant No. 272-08-0285 and the Discovery Center.

\appendix
\section{Derivation of Eq. (\ref{Ha}) }\label{apHa}

The growth factor satisfies the following differential equation:
\be
\delta''(a)+\left(\frac{3}{a}+\frac{H'(a)}{H(a)}\right)\delta'(a)
-\frac{3}{2}\frac{\omms \Geff(a)}{a^5 H(a)^2/H_0^2}\delta(a)=0
\label{Aode}\ee where $H(a)\equiv\frac{\dot{a}}{a}$ is the Hubble parameter and
Eq.~(\ref{Aode}) has the initial conditions $\delta(0)=0$ and $\delta'(0)=1$ for
the growing mode.

Eq.~(\ref{Aode}) can be rewritten in the form \cite{Starobinsky:1998fr},\cite{Nesseris:2007pa}: \be \left(\frac{H(a)^2}{H_0^2}\right)'+\frac{6}{a} \frac{H(a)^2}{H_0^2}=\frac{3 \omms  \Geff(a) \delta(a)}{a^5 \delta'(a)}-\frac{2 H(a)^2}{H_0^2 }\frac{\delta''(a)}{\delta '(a)} \label{Aode1}\ee

Solving Eq.~(\ref{Aode1}) for $H(a)^2$ we get \be \left[\frac{H(a)^2}{H_0^2} a^6 \delta'(a)^2\right]_{0}^{a}= 3\omms \int^a_0 x ~ \Geff(x) \delta(x)\delta'(x)dx \label{AHa} \ee Any realistic model should satisfy $\lim_{a \to 0}(\delta'(a)) \to 1$ and $\lim_{a \to 0}(H(a)^2 a^6)\to 0$ as $H(a)^2 \simeq H_0^2 \omms a^{-3}$ for $a\sim0$. Therefore, the LHS of Eq.~(\ref{AHa}) is equal to $H(a)^2 a^6 \delta'(a)^2$ and we get Eq.~(\ref{Ha}).

\section{The growth factor in a matter dominated Universe with $\Geff(a)$}\label{formulas}
In a matter dominated Universe ($\omms=1$ and $H(a)^2=H_0^2a^{-3}$) Eq.~(\ref{ode}) can be solved analytically for various cases for $\Geff(a)$. More specifically, we will consider the case \be \Geff(a)=g_0+\frac{g_2}{2} (a-1)^2\ee and we will show that only when $g_0=1-g_2/2$ the growth factor grows as $\delta(a)\simeq a$ for $a\ll 1$. In this case, Eq.~(\ref{ode}) can be solved and the growing mode for the growth factor $\delta(a)$ can be found to be: \ba \delta(a)&=&a^{-\frac{1}{4} +\frac{c_1}{2}} e^{-\frac{1}{2} \sqrt{3~g_2} a }  {}_1F_1(c_2;1+c_1;\sqrt{3g_2} a) \nn\\
c_2&=&\frac{1}{2} \left(-1+\sqrt{3~g_2}-c_1\right)\nn\\ c_1&=&\frac{1}{2} \sqrt{1+24 g_0+12 g_2}\nn\ea where ${}_1F_1(a;c;z)$ is a confluent hypergeometric function given as the limit ${}_1F_1(a;c,z)=\lim_{b\to\infty} {}_2F_1(a,b;c;z/b)$, see Ref.~\cite{handbook} for more details.

As a self-consistency test one can check that for $a\ll 1$, the growth factor scales as
\ba \delta(a) &\simeq& a \textrm{ for } g_0=1-g_2/2 \label{GGR}\\
\delta(a)&\simeq& a^{\frac{1}{4}\left(-1+\sqrt{25+12g_2}\right)} \textrm{ for } g_0=1 \label{Gmog}\ea where (\ref{GGR}) corresponds to a model that does not depart from GR at $a\ll 1$ while (\ref{Gmog}) corresponds to a model that has $\Geff(0)=1+\frac{g_2}{2}$.

\section{Notes on the error estimation for Tables III and IV.}\label{errap}
In this section we present some details about the jack-knife and the variance between several simulations approach.

The jack-knife is a non-parametric estimation of the variance or more general measures of the error, see Ref. \cite{efron}. Assume that we have a data set of size $n$ that consists of independent and identically distributed data points, \be X_1, ~ X_2, ~...~, ~X_n \ee and an estimator $\hat{\theta}=\hat{\theta}(X_1, ~ X_2, ~...~, ~X_n)$. Then assume that $\hat{\theta}_{(i)}=\hat{\theta}(X_1, ~ X_2, ~...~,~X_{i-1},~X_{i+1}, ~X_n)$ is the estimate after having deleted the ith point and $\hat{\theta}_{(.)}=\sum_{i=1}^n \hat{\theta}_{(i)}/n$, the average of the estimates $\hat{\theta}_{(i)}$. Then, the jack-knife estimate of the standard deviation is \cite{efron} \be \sigma_{jack}^2 =\frac{n-1}{n} \sum_{i=1}^n \left(\hat{\theta}_{(i)} -\hat{\theta}_{(.)}\right)^2 \label{sjack} \ee One caveat of the jack-knife estimate is that it fails for non-smooth statistics, for example while it works very well for the mean, it fails for the median (see Ref. \cite{efron}). As seen in Table \ref{table3}, this happens in our case as well, as our estimators Eqs. (\ref{omGR}), (\ref{omreal}) and (\ref{g2}) are not smooth.

For the second approach, we created synthetic data sets by adding gaussian noise on the WiggleZ data, ie $mock_i=(z_i,\fs_i,\sigma_i)$, where $\fs_i=\fs_{WiggleZ, i}+N_i$ and the $(z_i,\sigma_i)$ correspond to the real data. Then we calculate the value of each estimator of Eqs. (\ref{omGR}), (\ref{omreal}) and (\ref{g2}) and the error corresponds to the standard deviation of each sample, see Chapter 15.6 of Ref. \cite{Press:1992zz} for more details. This is different from the simulations we performed in Section IV.A, as that assumed a specific model for the synthetic data.

\end{document}